\documentclass[journal,12pt,onecolumn,draftclsnofoot,]{IEEEtran}
\usepackage{amsmath,amsfonts}
\usepackage{algorithmic}
\usepackage{algorithm}
\usepackage{array}
\usepackage[caption=false,font=normalsize,labelfont=sf,textfont=sf]{subfig}
\usepackage{textcomp}
\usepackage{stfloats}
\usepackage{url}
\usepackage{verbatim}
\usepackage{graphicx}
\usepackage{hyperref}
\usepackage{cite}
\usepackage{xcolor}
\usepackage{color}
\begin{document}

\title{A hollow-core fiber based stand-alone multimodal (2-photon, 3-photon, SHG, THG) nonlinear flexible imaging endoscope}

\author{D. Septier, G. Brévalle-Wasilewski, E. Lefebvre, N. Gajendra Kumar, Y. J. Wang, A. Kaszas, H. Rigneault, and A. Kudlinski

\thanks{Manuscript received xxx; revised xxx; accepted xxx. Date of publication xxx; date of current version xxx. }
\thanks{D. Septier was with Univ. Lille, CNRS, UMR 8523 - PhLAM - Physique des Lasers Atomes et Molécules, Lille, France. He is now with Lightcore Technologies, Marseille, France. (\href{dylan.septier@lightcore.tech}{\textcolor{blue}{dylan.septier@lightcore.tech}})}
\thanks{G. Brévalle-Wasilewski is with Lightcore Technologies, Marseille, France.} 
\thanks{E. Lefebvre is with Univ. Lille, CNRS, UMR 8523 - PhLAM - Physique des Lasers Atomes et Molécules, Lille, France and with Lightcore Technologies, Marseille, France. } 
\thanks{N. Gajendra was with Aix Marseille Univ, CNRS, Centrale Med, Institut Fresnel, Marseille, France. He is now with Lightcore Technologies, Marseille, France. } 
\thanks{Y. J. Wang is with Lightcore Technologies, Marseille, France. } 
\thanks{A. Kaszas is with Lightcore Technologies and the Multimodal Neurotechnology Group from the Institute of Cognitive Neuroscience and Psychology at the HUN-REN Research Centre for Natural Sciences (Hungary)} 
\thanks{H. Rigneault is with Aix Marseille Univ, CNRS, Centrale Med, Institut Fresnel, Marseille, France. (\href{herve.rigneault@fresnel.fr}{\textcolor{blue}{herve.rigneault@fresnel.fr}})} 
\thanks{A. Kudlinski is with with Univ. Lille, CNRS, UMR 8523 - PhLAM - Physique des Lasers Atomes et Molécules, Lille, France. (\href{alexandre.kudlinski@univ-lille.fr}{\textcolor{blue}{alexandre.kudlinski@univ-lille.fr}})} 
\thanks{The reported studies have been conducted following the EU ethical rules on animal. Mouse brain slices are coming from the Institut des Neurosciences de la Méditerranée following INMED agreement number B13-055-19 and OGM agreement number 5817.}}



\maketitle

\begin{abstract}
Multimodal nonlinear endoscopes have been a topic of intense research over the past two decades, enabling sub-cellular and label-free imaging in areas not reachable with table-top microscopes. They are sophisticated systems that can be implemented on an optical table in a lab environment, but they cannot be easily moved within or out of the lab. We present here a multimodal and flexible nonlinear endoscope system able to perform two photon excited fluorescence and second harmonic generation imaging with a stand-alone and moveable kart integrating a compact ultrashort laser source. In addition, the system can perform three photon excited fluorescence and third harmonic generation thanks to a delivery optical fiber used to deliver ultrashort pulses from massive and not movable laser systems into the stand-alone kart. The endoscopic fiber probes and delivery optical fibers are based on functionalized negative curvature hollow core fibers. The endoscope distal head has a diameter $<$2.2mm and can perform nonlinear imaging at max 10 frames/s over a field of view up to 600 µm with a $\sim$1 µm spatial resolution.

\end{abstract}


\section{Introduction}
\IEEEPARstart{M}{ultiphoton} microscopy \cite{zipfel_live_2003,sheppard_multiphoton_2020,carriles_invited_2009} has now matured and is widely recognized as an essential technology for applications in biological and medical sciences. The appeal of multiphoton microscopy lies in two key factors. Firstly, it offers a variety of nonlinear contrast mechanisms activated by short laser pulses with near-infrared wavelengths, enabling penetration into biological tissues \cite{miller_deep_2017}. Some contrast mechanisms, like two-photon excited fluorescence (2-photon) \cite{diaspro_multi-photon_2006}, require fluorescent molecules, while others, termed 'label-free' contrasts, can be activated without any labeling or staining. These include two-photon excited auto-fluorescence \cite{ranawat_recent_2019}, second (SHG) \cite{aghigh_second_2023} and third (THG) \cite{weigelin_third_2016} harmonic generation, four-wave mixing (FWM) \cite{min_near-degenerate_2009}, such as chemical-sensitive coherent anti-Stokes Raman scattering (CARS) and stimulated Raman scattering (SRS) \cite{min_coherent_2011,rigneault_tutorial_2018} along with various pump-probe schemes \cite{fischer_invited_2016} addressing transient absorption or vibrations lasting longer than the excitation pulse. Tremendous applications have been demonstrated in many fields of biology \cite{ishii_focusing_2022,konig_multiphoton_2000} but also in medical sciences \cite{wang_two-photon_2010,orringer_rapid_2017}.
Besides its successful implementation in microscopy \cite{yue_multimodal_2011}, there have been numerous attempts to implement multiphoton imaging in flexible probes that can be handheld or inserted in constrained environments, such as those found in endoscopy for the exploration of the human body. Multiphoton endoscopes have been reported using MEMS \cite{tang_design_2009} or resonant piezo-tubes \cite{myaing_fiber-optic_2006,do_fiber-optic_2014,rivera_use_2012} to build tiny scanners that can be located at the endoscope distal head. Imaging fiber bundles have also been used to move the scanning procedure away from the sample side, this remove all moving parts or driving currents from the distal endoscope head \cite{lukic_endoscopic_2017}. The main challenge of multiphoton endoscope is to deliver ultra-short pulses at the sample plane to activate the nonlinear contrasts and several solutions have been reported such as pre-compensation schemes \cite{wu_scanning_2009,murari_compensation-free_2011,ducourthial_development_2015} or using hollow core fibers \cite{lombardini_high-resolution_2018,kudlinski_double_2020}. With these advances multiple flexible multimodal nonlinear endoscopy has been reported using one or a combination of the 2-photon, SHG, THG and CARS contrast mechanisms \cite{myaing_fiber-optic_2006,lombardini_high-resolution_2018,pshenay-severin_multimodal_2021,septier_label-free_2022,kim_lissajous_2019,zhang_compact_2012,zhao_development_2010,akhoundi_compact_2018,fu_nonlinear_2006,chang_two-photon_2008,rivera_compact_2011,li_twist-free_2021} and a recent review can be found in \cite{kucikas_two-photon_2023}. So far three photon excited fluorescence (3-photon) flexible endoscope reports are more sparse \cite{septier_label-free_2022} and most of the current 3-photon implementation are using rigid graded index (GRIN) lenses \cite{klioutchnikov_three-photon_2020,huland_three-photon_2013}.
Most implementations of multiphoton flexible endoscopes are lab-based systems that cannot be easily moved within the lab space or to an external field where measurements need to be performed. In this paper, we overcome this limitation and present a stand-alone integrated flexible multiphoton endoscope system placed on a movable rack platform. This autonomous system includes a small frame femtosecond (fs) laser system but can also be connected through a delivery fiber to a larger frame multiphoton fs or picosecond (ps) laser platform to extend multiphoton contrasts. The system is controlled by dedicated user-friendly software and can perform 2-photon and SHG using in in-built laser as well as 3-photon and THG using the fiber delivery from a larger laser frame. The endoscope can perform imaging over field of view (FoV) up to 600 µm with a resolution of $\sim$ 1 µm at a maximum frame rate of 10 frames/s (fps).

\section{Double-clad hollow core fiber and functionalization of the fiber tip}

The nonlinear endoscope system developped is based on double clad hollow core negative curvature optical fibers similar to the ones presented in details in \cite{kudlinski_double_2020,septier_label-free_2022}. They are indeed excellent candidates to deliver ultrashort pulses thanks to their intrinsic low group velocity dispersion (GVD) and nonlinearity \cite{kolyadin_negative_2015,wei_negative_2017,debord_ultralow_2017}. However, they need to be functionalized in order to strongly reduce the output spot size, which is required for high resolution imaging \cite{lombardini_high-resolution_2018,septier_label-free_2022}. We are addressing all these issues in this section. 

\subsection{Double-clad anti-resonant hollow core fiber}

\begin{figure}[!h]
    \centering
    \includegraphics[width=0.5\linewidth]{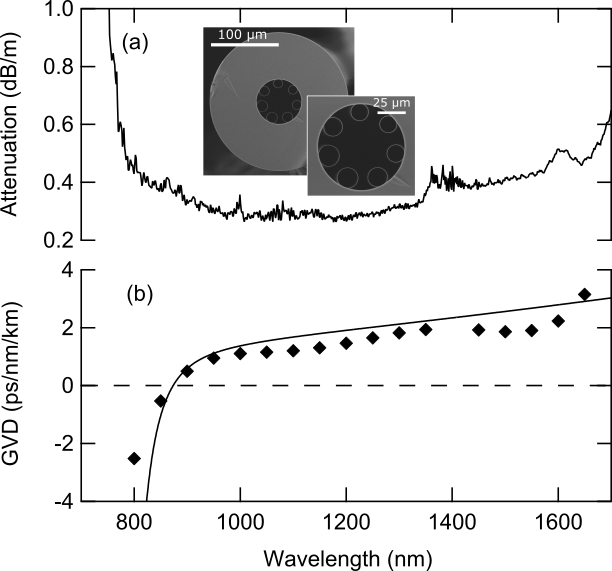}
    \caption{Hollow core fiber: (a) Measured attenuation spectrum. Inset: SEM images. (b) Measured GVD (markers) and simulation (solid line) using the parameters in the text and the model from \cite{zeisberger_analytic_2017}.}
    \label{fig:fibre}
\end{figure}

The inset of Fig. \ref{fig:fibre}(a) shows a scanning electron microscope (SEM) image of the fiber used to build the flexible endoscope, with a close-up on the hollow core region. The design is made of 7 non-touching capillaries of 14 µm average external diameter, 8 µm average separation and 350 nm average thickness. The core radius is 18 µm. The silica cladding surrounding the central hollow region is able to guide light \textit{via} total internal reflection thanks to a low-index polymer coating (not shown on the picture). The numerical aperture (NA) of the double clad was measured to be 0.38 at 450 nm \cite{kudlinski_double_2020}. 
Fig. \ref{fig:fibre}(a) shows the attenuation spectrum obtained from a cutback measurement from 17 m to 1.5 m in the first antiresonant window. The attenuation is 0.32 dB/m at 920 nm and 0.31 dB/m at 1300 nm, which are the two wavelengths of main interest here. The fiber was found to be single mode for fibers longer than 2 m. The $1/e^2$ mode field diameter (MFD) was measured to be equal to 24 µm at 920 nm. Fig. \ref{fig:fibre}(b) shows the measured and simulated GVD, respectively displayed with markers and solid line. The measurement was done using a low coherence Mach-Zehnder interferometer \cite{tateda_interferometric_1981} and the simulation was performed with the analytical model \cite{zeisberger_analytic_2017} using the geometrical parameters listed above. The GVD value is 0.85 and 2.1 ps/nm/km at 920 and 1300 nm, respectively.

\subsection{Functionalization of the fiber tip}

Since the MFD of the hollow core fiber is as large at 24 µm, it needs to be strongly reduced for endoscopy application that require a $\sim$1 µm spot size at the sample plane. This can be done for example by splicing silica beads \cite{lombardini_high-resolution_2018,kudlinski_double_2020} or short pieces of graded-index (GRIN) fibers \cite{septier_label-free_2022} at the output facet. We chose this last solution because in addition to focusing the output beam, it allows to seal the fiber end-face and brings more mechanical robustness than the bead splicing method. For a given GRIN fiber, the smallest focal spot is obtained when the fiber length is exactly half the self-imaging period \cite{iga_theory_1980}. In this case, the focal spot is located on the GRIN fiber end-surface. This is fine for the pulse duration around $\sim$150 fs and pulse energy in the $\sim$nJ range required for 2-photon absorption. However, 3-photon absorption requires significantly shorter pulses ($<$100 fs) with higher energy ($\sim$µJ), such that a parasitic THG signal can be generated at the interface between the GRIN fiber end-face and air \cite{septier_label-free_2022}. In this case, the GRIN fiber can be shortened so that the focal spot is located further away in air, at the expense of a less efficient MFD reduction (\textit{i.e.} focalization). This shortening reduces the energy density at the GRIN fiber/air interface and therefore strongly reduces the parasitic THG signal. In the present work, we present two different fiber probes: the first one (labeled 2-photon fiber) is optimized for 2-photon absorption and features a GRIN fiber length that performs the smallest MFD at the GRIN fiber/air surface and the second one (labeled 3-photon fiber) has a GRIN fiber length optimized for 3-photon absorption and performs a smaller MFD reduction but is free of THG noise.
\begin{figure}
    \centering
    \includegraphics[width=0.5\linewidth]{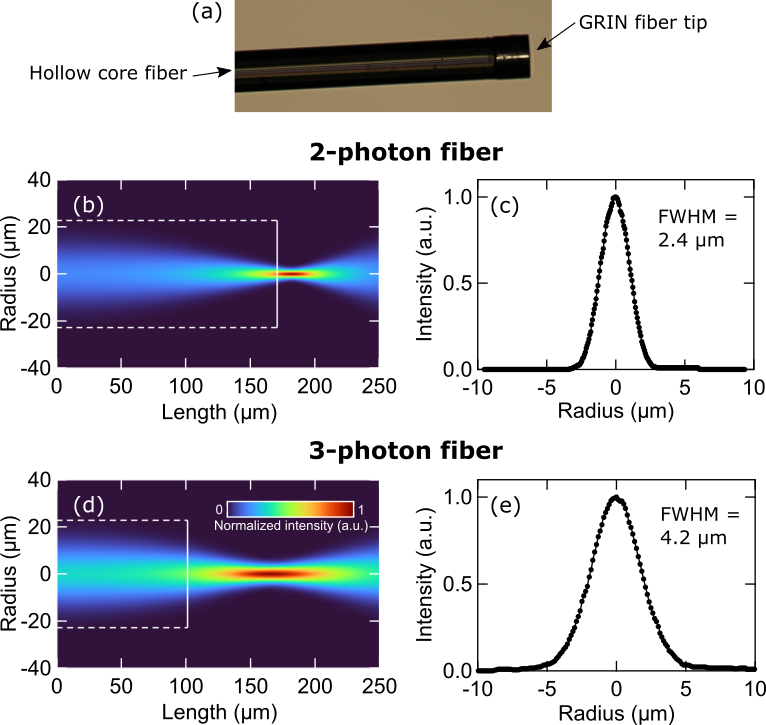}
    \caption{(a) Optical microscope image of a hollow core fiber functionalized with a GRIN fiber. (b) Simulation of the optical intensity evolution at 920 nm for a GRIN fiber length of 171 µm. White dashed lines represent the core/cladding interface of the GRIN fiber and the solid white line is the GRIN fiber end-face. (c) Measured spot profile at focus at 920 nm. (d) Simulation of the optical intensity evolution at 1300 nm for a GRIN fiber length of 101 µm and (e) measured spot profile at focus at 1300 nm (in air, 64 µm after the GRIN fiber end-face). }
    \label{fig:simu}
\end{figure}

Fig. \ref{fig:simu}(a) shows an optical microscope image of the 2-photon fiber that has a diameter of 200 µm. The double-clad hollow core fiber is on the left and the GRIN fiber which has been spliced is located on the right. The GRIN fiber has a core diameter of 48 µm and a parabolic refractive index profile with a maximum refractive index difference of 30$\times$10$^{-3}$ between the core center and the cladding. The length of the GRIN fiber segment was optimized through numerical simulations to obtained the smallest achievable MFD at its end-face (\textit{i.e.} smallest focal spot). Numerical simulations were performed using the model of [46]. Fig. \ref{fig:simu}(b) shows the simulated beam evolution in a 171 µm long GRIN fiber (as measured from the experimental fiber of Fig. \ref{fig:simu}(a)). The input beam corresponds to the fundamental mode of the hollow-core fiber (24 µm MFD), white dashed lines represent the core/cladding interface of the GRIN fiber and the solid white line is the GRIN fiber end-face. The simulated focal spot has a full width at half maximum (FWHM) of 2.37 µm, located 11 µm away in the air from the GRIN fiber end-face. This is because the theoretical half self-imaging period is 188 µm and the actual GRIN fiber is slightly shorter (171 µm) due to fabrication uncertainties. The intensity profile measured at focus is displayed in Fig. \ref{fig:simu}(c). It has a FWHM of 2.4 µm, in excellent agreement with simulations. 
The 3-photon fiber was designed with a much shorter GRIN fiber tip in order to address the parasitic THG noise issue raised above. Fig. \ref{fig:simu}(d) shows the beam evolution versus propagation distance for a GRIN fiber length of 101 µm. The focal spot is located 64 µm away from the fiber end-face, which significantly reduces the energy density on the fiber/air interface and the resulting THG. The simulated FWHM of the focused beam is 4.32 µm, in excellent agreement with the measured one of 4.2 µm, displayed in Fig. \ref{fig:simu}(e).

To summarize, the 2-photon fiber has a 2.4 µm FWHM beam located almost on the GRIN fiber end face and the 3-photon fiber has a larger beam size at focus of 4.2 µm, located 64 µm away from the GRIN fiber end-face.

\subsection{Distal head}

\begin{figure}
    \centering
    \includegraphics[width=0.5\linewidth]{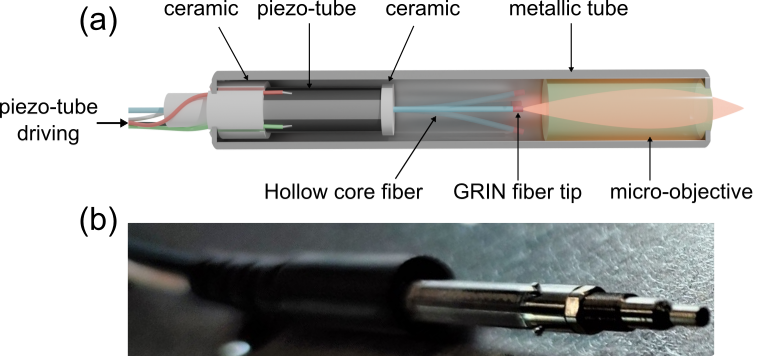}
    \caption{(a) Scheme of the inside of the distal endoscopic head. Distal head dimensions: 1.3 mm to 2.2 mm in diameter (depending on the micro-objective) and 40 mm in length. Total weight $<$2 g. (b) Picture of the distal head. }
    \label{fig:head}
\end{figure}

Fig. \ref{fig:head}(a) shows a sketch of the inside of the distal head. The uncoated end of the hollow core fiber (in blue) is glued into a four quadrant piezo-tube in order to perform a scan. The piezo-tube quadrants are driven by four electrical connections (see details in next Section) and are hold into the endoscope metallic tube by two ceramics. The laser beam focused spot formed at the end of the GRIN fiber piece (in red) is re-imaged on the sample with a miniature micro-objective that is integrated in the surrounding metallic tube.  We used either a custom micro-objective based on four commercially available achromatic doublets (labeled 4L-MO) or a commercial GRIN micro-objective (GRIN-MO). Their full optical properties as well as the detailed characteristics of the output spot obtained with both fibers is detailed in Section V hereafter. Fig. \ref{fig:head}(b) is a photograph of the manufactured distal head that has a diameter of 1.3 mm (with the 4L-MO) and 2.2 mm (with the GRIN-MO) and a length of 40 mm. It total weight is $<$2 g. The proximal side of the endoscope is mounted with a FC/PC connector (not shown) and the length of the endoscope fiber is typically 2.5 m. The transmission of the full endoscope was measured to be above 40\% in the 920 nm - 1300 nm range. 

\section{Integrated resonant piezo scanner and spiral scanning image acquisition}

The functionalized fiber presented in the previous sections is inserted into a piezoelectric tube (Fig. \ref{fig:head}) in order to make a spiral scanning trajectory \cite{myaing_fiber-optic_2006,do_fiber-optic_2014,ducourthial_development_2015,lombardini_high-resolution_2018,septier_label-free_2022,rivera_compact_2011}. In this section, we present technical details we have implemented in order to improve the image acquisition. 

\subsection{Integrated resonant piezo scanner}
\begin{figure}[!b]
    \centering
    \includegraphics[width=0.5\linewidth]{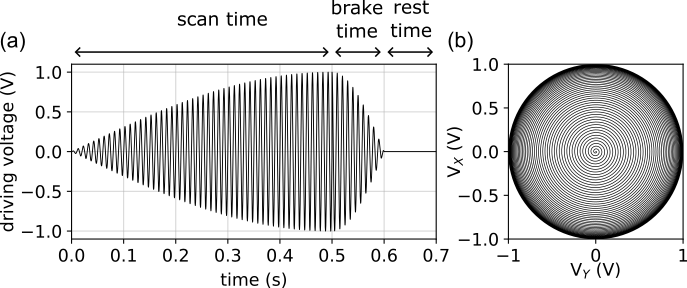}
    \caption{(a) Excitation signal of one axis corresponding to a full image cycle, including scanning, braking and rest time. (b) Spiral pattern produced by the drive signal during the scan phase.}
    \label{fig:scan}
\end{figure}
Fig. \ref{fig:scan}(a) is an illustration of the signal used to drive the 4 quadrants piezoelectric tube (PT Piezo scanner tubes, PI) in order to obtain a spiral scan. Only the drive voltage along one direction is shown in Fig. \ref{fig:scan}(a). In the orthogonal direction the drive signal is similar, but with a $\pi/2$ phase difference which produce a spiral scan as shown in Fig. \ref{fig:scan}(b). The scanned cycle is composed of a ‘scan time’, which is a sine wave along one axis and a cosine wave along the other axis, both at a frequency equal to the resonant frequency of the free-standing fiber that is attached to the 4-quadrant piezo tube (see Fig. \ref{fig:head}). An amplitude modulation is also added during this ‘scan time’ that expands the spiral radius over time (Fig. \ref{fig:scan}(a), scan time). This is during this ‘scan time’ that the optical signals (2-Photon, SHG, …) are acquired. The ‘scan time’ ends when the spiral reaches its maximum radius. Then starts the ‘brake time’ where and active braking is performed by keeping the resonant drive voltages along the two $(x,y)$ piezo axes but with opposite phases to the resonant fiber movement. The reason for this ‘brake time’ is to bring the fiber back to its original position, at the center, as quick as possible before the next ‘scan time’ starts and acquires the next image. An optional ‘rest time’ can also be added to ensure that no residual oscillation is present (Fig. \ref{fig:scan}(a)).

\subsection{Pixel dwell time}

\begin{figure}
    \centering
    \includegraphics[width=0.5\linewidth]{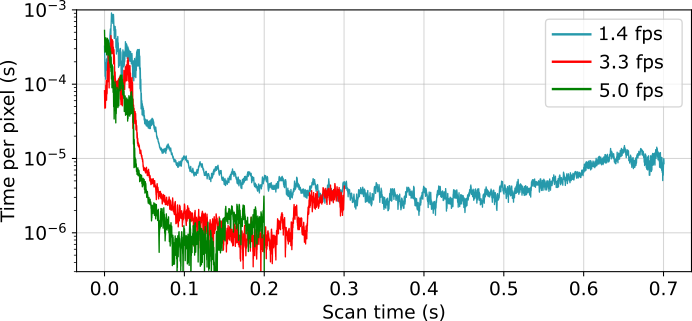}
    \caption{Pixel dwell time during the scan phase, calculated from experimental $x(t)$ and $y(t)$ positions, for different acquisition rates of 1.4 fps (blue line), 3.3 fps (red line) and 5 fps (green line). }
    \label{fig:dwell}
\end{figure}

The spiral scan (‘scan time’, 'brake time', 'rest time') requires a proper calibration prior to imaging that is performed in a dedicated calibration step described below. This calibration step is robust enough to be valid during a full day of experiment and has been made automatic to facilitate the operation of the endoscope. We describe below the calibration of the 'scan time' which is the most relevant to perform imaging. During the calibration a position sensor detector (PSD) provides the output position ($x$ and $y$) of the focused spot at the endoscope image plane as a function of time during the ‘scan time’. The known $x(t)$ and $y(t)$ positions are necessary to reconstruct the final image by matching the recorded optical signal (2-photon, SHG,...) with the sample spot position $x(t)$ and $y(t)$. Because the scanner is working at resonance, the time to achieve a circle within the spiral scan is the same whatever the radius in the spiral. This leads to an uneven sampling grid where the center of the image sees the laser beam much longer than the outer part of the FoV. To get some insight into this peculiar scanning scheme we can count the number of data acquisitions corresponding to each image pixel across the FoV. Moreover, we can follow the spiral pattern and match each pixel with the number of acquisitions it contains, thus providing the time spent on each pixel (the pixel dwell time) as a function of time. This is illustrated in Fig. \ref{fig:dwell}. The pixel dwell time is calculated for different acquisition rates of 1.4 fps (blue line), 3.3 fps (red line) and 5 fps (green line). For each acquisition rate investigated, the pixel dwell time is at least two orders of magnitude larger at the beginning of the scan (at the center of the image) as compared to the remaining pixels at latter time. This phenomenon causes photobleaching issues at the center of the image where the sample is over illuminated. This problem is solved in our system using a dedicated power management across the scan.

\subsection{Power management}

\begin{figure}
    \centering
    \includegraphics[width=0.5\linewidth]{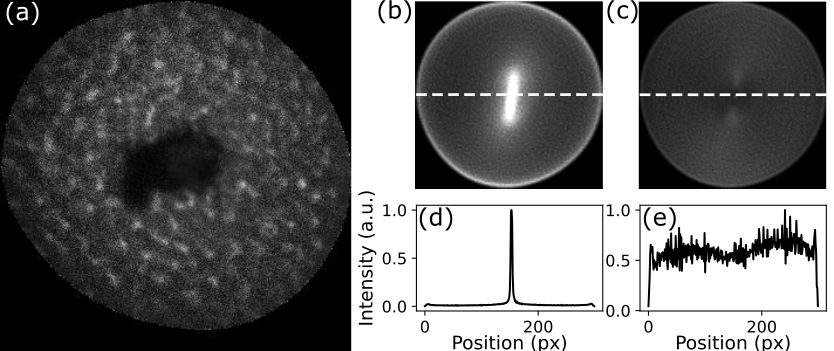}
    \caption{(a) 3-photon image of a fixed mouse retina sample, showing an example of photobleaching spot at its center. Image of fluoresceine liquid film reconstructed from a typical experimental calibration for (b) a constant signal and (c) a signal adjusted using the time per pixel calculated beforehand. (d) and (e) are intensity plots along dashed lines of (b) and (c) respectively. 1 fps, 100 µm FoV, micro-objective: 4L-MO}
    \label{fig:power}
\end{figure}

The high power density at the center of the image is inherent to the spiral shape of the resonant scan, and can lead to photobleaching or sample burning, as explained above. An example of photobleaching observed while imaging a fixed retina tissue is exhibited in Fig. \ref{fig:power}(a). This can also be evidenced by reconstructing an image with a constant signal (fluoresceine liquid layer) where the center appears very bright (Fig. \ref{fig:power}(b)). This image is voluntarily overexposed to allow the weaker signal away from the center to be visible. The plot in Fig. \ref{fig:power}(d) is the intensity profile along the dashed line of Fig. \ref{fig:power}(b). It shows a very steep peak at the center, emphasizing further this issue. To overcome this problem, we can divide the signal measured by the dwell time calculated above. Applying this to case of Fig. \ref{fig:power}(b) leads to the image reconstructed in Fig. \ref{fig:power}(c). The plot in Fig \ref{fig:power}(e), which is the intensity profile along the dashed line of Fig. \ref{fig:power}(c), features an almost flat signal in good agreement with the fluoresceine concentration in this test sample. This re-scaling does not avoid the photobleaching at the image center. To solve this problem we have added an acousto-optic modulator (AOM) before injection into the endoscope which controls the excitation laser power over time. By controlling the ramp-up power over time we can make the average power delivered per pixel constant, with this the spiral illumination scheme is not harmful to the sample at the center of the FoV.

\subsection{Image reconstruction}

In practice, we found that it is almost impossible for the fiber to be completely immobile after a full image cycle. Some residual movements are always present and affect the subsequent cycle. The result is an uncertainty of the exact position of the fiber at a given time during a multiple image scan. This leads to a visible movement between frame-to-frame images, even though the object imaged is actually static. Such an effect is usually minimal, to the point of being almost invisible, but it can visually affect the images for high repetition rates ($>$5Hz), as evidenced by the animation of \textcolor{blue}{Supp info movie 1} (left). To circumvent this effect, we developed a motion compensation algorithm based on the opensource module open-cv and using the so-called Demon's algorithm \cite{thirion_image_1998}. The real-time algorithm can detect these parasitic movements and compensate for them, as showed in the reconstructed video of \textcolor{blue}{Supp info movie 1} (right). 

\subsection{Control software}
All parameters related to scanner parameters and calibration, laser power, as well as image acquisition and processing are controlled from a user-friendly control software. Details and examples of screenshots are provided in Appendix A. This makes the endoscope system easy to use to a non-specialist user and provides stacks of images that can be analyzed with conventional softwares (ImageJ, Suite2P, etc) but also used in user code (MatLab, Python, etc). 

\section{Integrated stand-alone endoscope rack}

\begin{figure}
    \centering
    \includegraphics[width=0.5\linewidth]{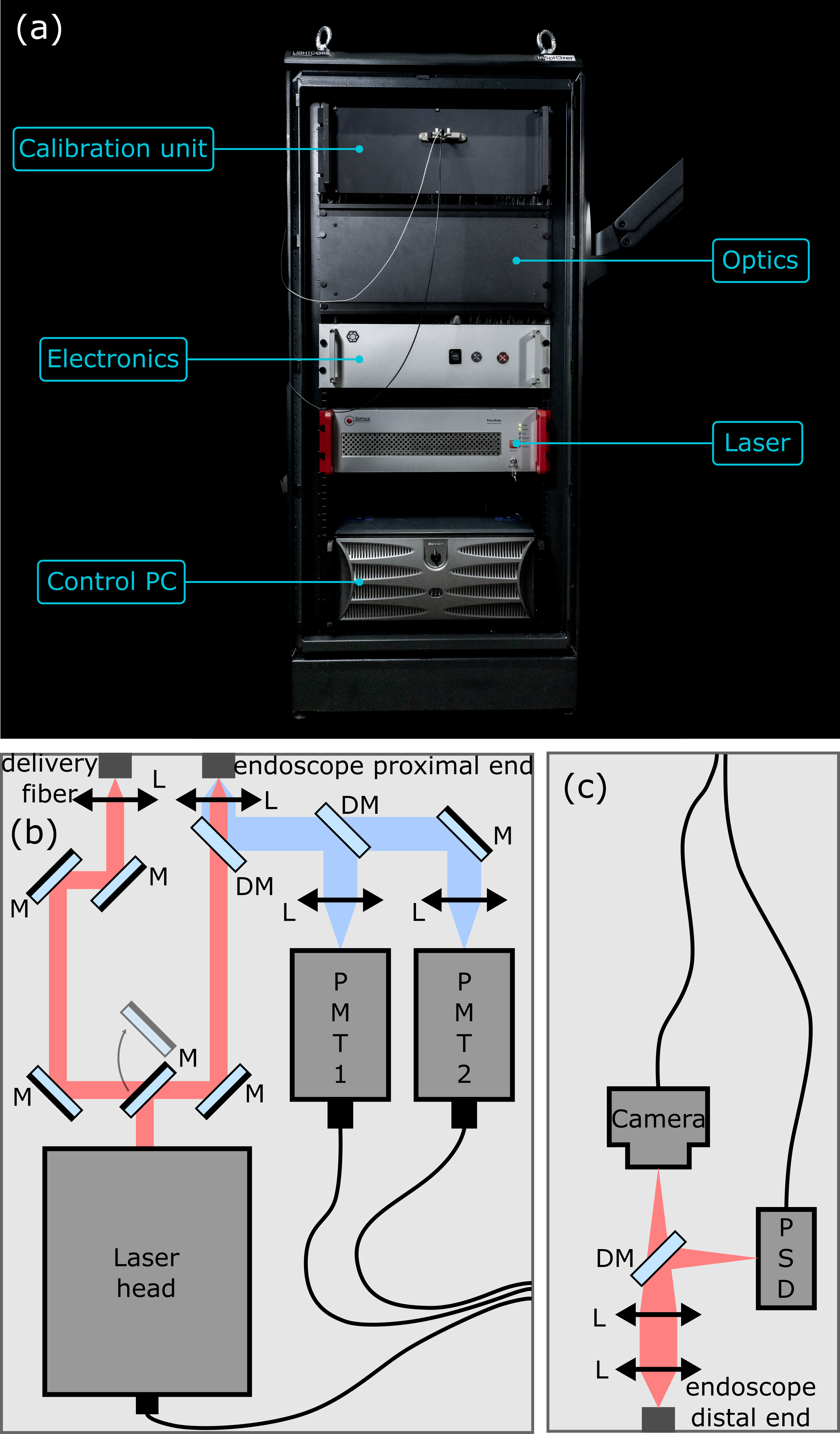}
    \caption{(a) Picture of the endoscope integrated cabinet. (b) Scheme of the optical 'coupling/detection unit'. The excitation laser can be a laser head integrated into the endoscope kart or coming from a delivery fiber (see section VII). (c) Scheme of the 'calibration unit'. L: lens ; M: mirror ; DM: dichroïc mirror ; PMT: photomultiplier tube.}
    \label{fig:rack}
\end{figure}

The whole endoscope and its control units are racked in a movable cabinet as shown in Fig. \ref{fig:rack}(a). The stand-alone multiphoton endoscope system contains a compact and rackable femtosecond fiber laser. In the present study, it is a 920 nm, $>$1.5 W, 80 MHz, $<$100 fs pulse duration laser (Femto Fiber ultra 920, Toptica). The control electronics required to drive the piezo scanner and the control computer are also included in the cabinet. The monitor and keyboard are included \textit{via} a movable arm attached to the side. The cabinet also contains two optical units. The first one, represented in Fig. \ref{fig:rack}(b), contains a coupling stage required to inject light from the laser source into the endoscope hollow core fiber, either directly from the fiber laser head integrated into the cabinet, or from a laser located outside the rack using a dedicated hollow core delivery fiber (see section below). A flipping mirror located inside the unit allows to switch between these two configurations (internal or external lasers). The nonlinear signal (2-photon, SHG,…) collected from the sample is separated from the excitation laser by a dichroic mirror, and two photomultiplier detectors can be placed to simultaneously detect two nonlinear signals depending on the filters applied.

The second optical unit displayed in Fig. \ref{fig:rack}(c) is the 'calibration unit'. As mentioned before this is mandatory to record the spiral focus point trajectory of the ‘scan time’. For that the distal head of the endoscope is directly plugged to the front of the integrated cabinet and the moving focused spot at the endoscope distal side is imaged onto both a camera and a PSD. The camera is used to adjust the injection into the hollow core fiber and the PSD records the positions ($x(t)$,$y(t)$) of the spot during the scan. It allows to map the shape of the spiral scan and adjust the driving voltage parameters to obtain the most symmetric and extended spiral pattern that directly defined the endoscope FoV. This ($x(t)$,$y(t)$) calibration is required to reconstruct images during live acquisition as positions are attributed to each intensity measured, thus allowing to fill each pixel of the image with the appropriate measured intensity. Typically the endoscope can image a FoV of 600 µm at at maximum of 10 fps. The transmission of the full endoscope system is typically above 40\% that provides sufficient power to perform nonlinear imaging using a $>$1 W laser system. 

\section{Imaging performances}

The endoscope distal head (Fig. \ref{fig:head}) can be assembled either with the 2-photon or the 3-photon fiber described in section II B. In each case, two different micro-objectives can be used. The first one is a micro-objective based on four commercially available achromatic doublets (4L-MO) with a theoretical magnification of 0.63, a NA of 0.3 on the fiber side and 0.45 on the sample side and a working distance (WD) of 600 $\mu$m in air. The second one is a commercially available GRIN micro-objective (GRIN-MO) (GT-MO-070-016-ACR-VISNIR-30-20, GRINTECH) with a magnification of 0.22, a NA of 0.16 on the fiber side and 0.7 on the sample side and a WD of 230 $\mu$m in air. 

We measured the point spread function (PSF) and WD with calibrated optical setups for each fiber with both micro-objectives. Results are summarized in Table \ref{tab:table1}. 
\renewcommand{\arraystretch}{1.8}
\begin{table}[!t]
\caption{Summary of the optical performances of the 2-photon and 3-photon fibers with the two different micro-objectives.\label{tab:table1}}
\centering
\begin{tabular}{|c|c|c|c|c|}
\hline
MO  & \multicolumn{2}{|c|}{4L-MO} & \multicolumn{2}{|c|}{GRIN-MO}\\
\hline
  & PSF (µm) & WD (µm) & PSF (µm) & WD (µm)\\
\hline
2-p fiber (920 nm) & 1.58 & 600$\pm$15 & 0.92 & 220$\pm$15\\
\hline
3-p fiber (1300 nm) & 2.65 & 600$\pm$15 & 1.52 & 220$\pm$15\\
\hline
\end{tabular}
\end{table}
The choice between these two micro-objectives depends on the expected imaging performances. The 4L-MO has a larger WD of 600±15 µm so it can be used for deep imaging, at the expense of a lower PSF of 1.58 µm with the 2-photon fiber and 2.65 µm with the 3-photon fiber. The GRIN-MO allows much smaller PSFs of 0.92 µm and 1.52 µm respectively with the 2-photon and 3-photon fibers, but with a shorter WD of 220±15 µm. 


\section{Imaging with the built-in fs laser}

In order to test the imaging ability of our system, we performed a first round of experiments on various biological tissues using the fully integrated kart. The laser used is the 920 nm fs fiber laser although 780 nm (1 W) and 1050 nm (5 W) fs lasers are also available. This was done using the 2-photon fiber associated with either the 4L-MO or GRIN-MO. 

\subsection{Tissue samples}

\begin{figure}
    \centering
    \includegraphics[width=0.5\linewidth]{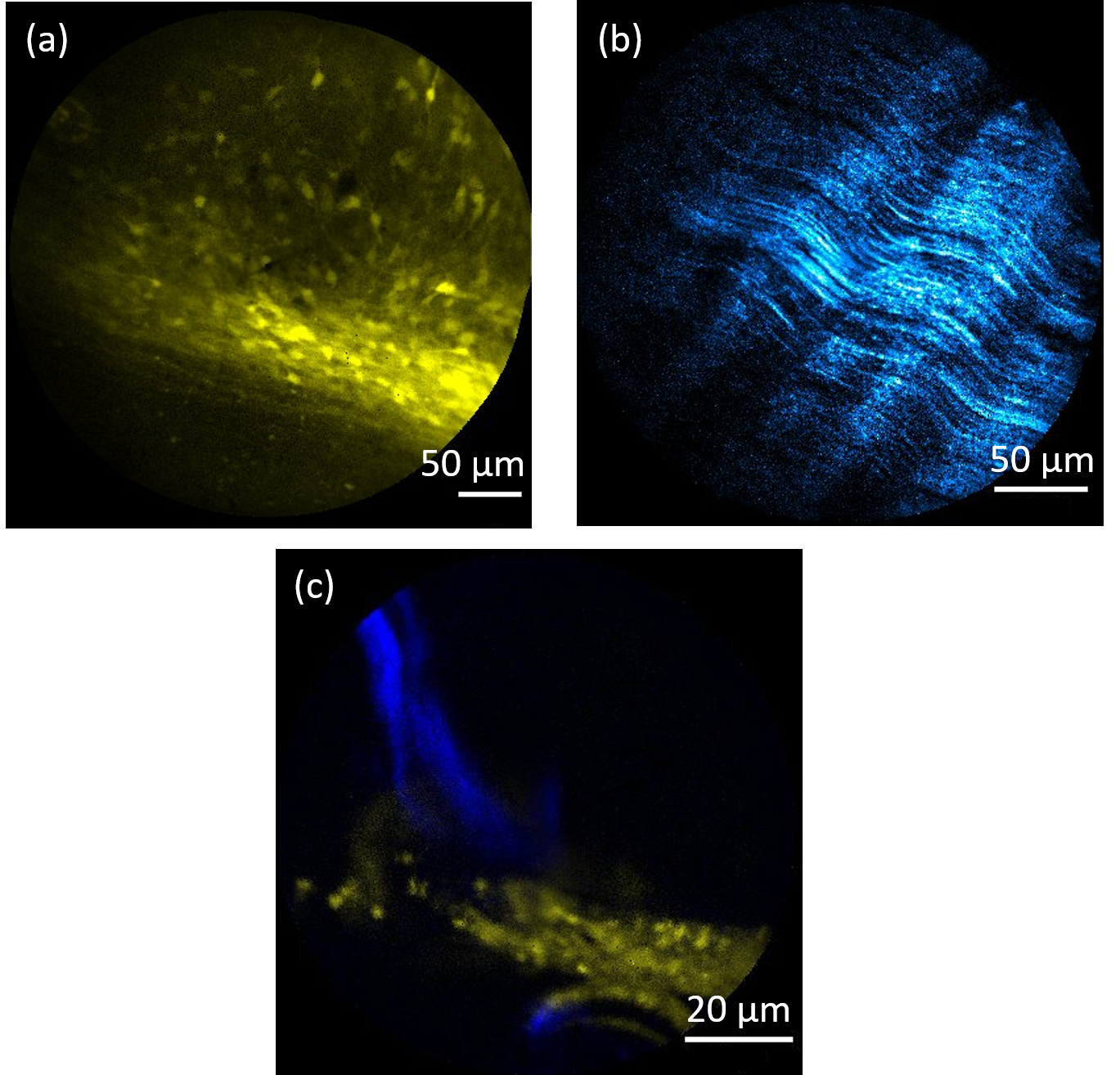}
    \caption{(a) 2-photon image of fixed mouse brain. (b) SHG image of yellow elastic connective tissue. (c) composite 2-photon (yellow) and SHG (blue) image of rat peritoneum. See acquisition details in the text.  }
    \label{fig:tissus}
\end{figure}

Fig. \ref{fig:tissus} shows a collection of images taken with the flexible endoscope system on different biological samples.  Fig. \ref{fig:tissus}(a) shows a 2-photon image taken on a fixed mouse brain labelled with GFP proteins (FoV 416 µm, 1 fps, average power at the sample 65 mW, 4L-MO). It highlights neurons in the hippocampus. Fig. \ref{fig:tissus}(b) is a SHG image of fixed yellow elastic connective tissue showing collagen fibrils structures (FoV 280 µm, 1.4 fps, 25 mW power, 4L-MO). Fig. \ref{fig:tissus}(c) is a multimodal 2-photon (yellow) and SHG (blue) image of fixed rat peritoneum showing the elastin and collagen (FoV 98 µm, 1 fps, 25mW power, GRIN-MO).

\subsection{Plant samples}

\begin{figure}[!hb]
    \centering
    \includegraphics[width=0.5\linewidth]{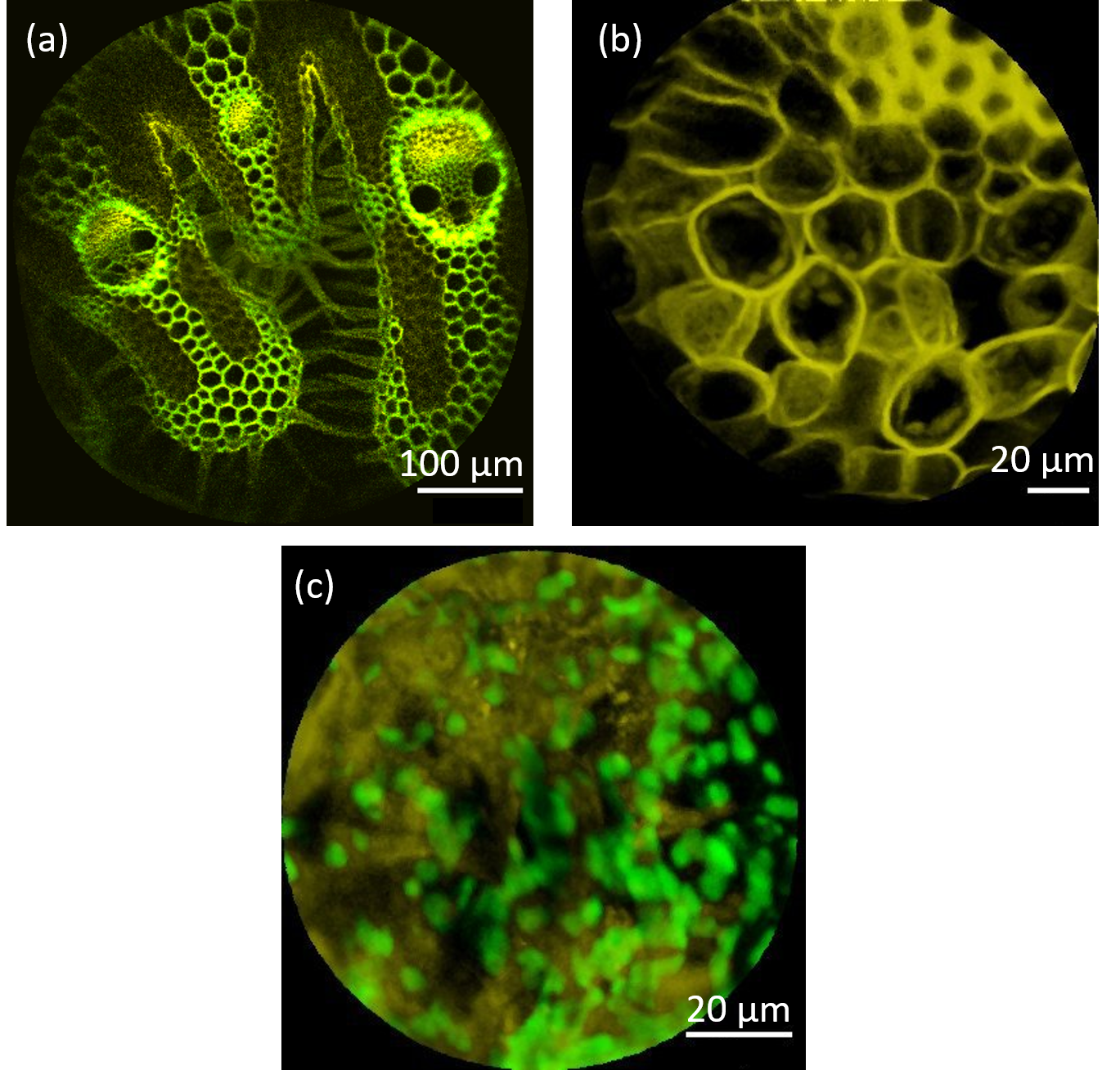}
    \caption{2-photon images of (a) stained marram grass (\textit{ammophila arenaria}) leaf at 520 nm (green) and 630 nm (yellow), (b) of stained holly (\textit{ilex aquifolium}) leaf and (c) 2-photon auto-fluorescence of fresh orchid leaf at 520 nm (yellow) and 630 nm (green). See acquisition details in the text.}
    \label{fig:plant}
\end{figure}
Next we tested the endoscope system for plant imaging, with some examples displayed in Fig. \ref{fig:plant}. Fig. \ref{fig:plant}(a) shows a 2-photon image of fixed marram grass (\textit{ammophila arenaria}) leaf stained with iodine green carmine stain (FoV 500 µm, 1.4 fps, power 20 mW, 4L-MO). Yellow parts correspond to the 2-photon fluorescence  filtered at 630$\pm$46 nm and green parts are taken at 520$\pm$30 nm. Folded structures used to trap water and vascular bundles typical of \textit{ammophila arenaria} leafs can be seen. Fig. \ref{fig:plant}(b) is a 2-photon image of holly (\textit{ilex aquifolium}) leaf stained with lugol's iodine (FoV of 168 µm, 1 fps, 20 mW, GRIN-MO, detection filter 520$\pm$30 nm). The system can also be used to study 2-photon auto-fluorescence and Fig. \ref{fig:plant}(c) is a 2-photon image of a fresh orchid leaf taken directly in a plant pot. Lignin is shown in yellow (acquired with a 520$\pm$30 nm filter) and chlorophyll in green with a 630$\pm$46 nm filter  (FoV of 98 µm, 1 fps, 20 mW, GRIN-MO).

\subsection{\textit{In vivo} imaging}

\begin{figure}[!h]
    \centering
    \includegraphics[width=0.3\linewidth]{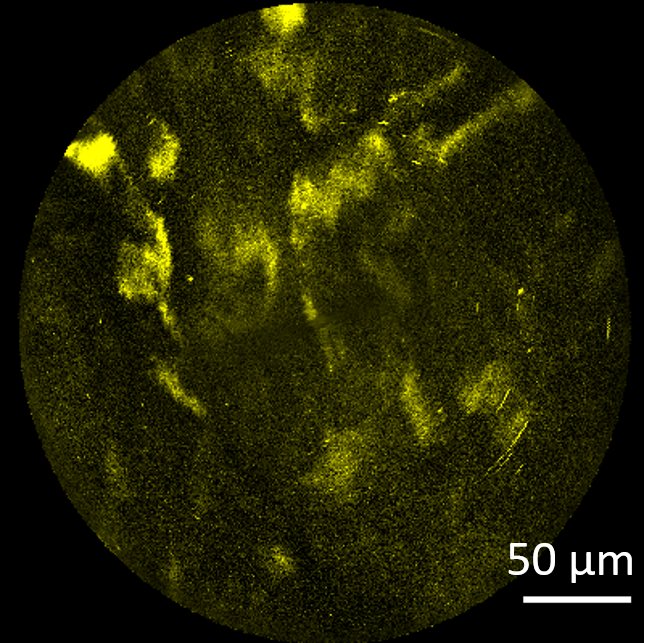}
    \caption{2-photon autofluorescence live imaging of human skin by holding manually the endoscope housed in the dedicated pen-like accessory directly onto the forearm. }
    \label{fig:skin}
\end{figure}

The manipulation of the distal head is easy to perform \textit{in vivo} imaging. As an example, Fig. \ref{fig:skin} shows a 2-photon auto-fluorescence image of human skin taken by manually holding the endoscope housed in a pen-like accessory in one hand and moving either its integrated miniature translation stage to find the signal, or sweeping the endoscope on the skin with the other hand (FoV 300 µm, 6 fps, 70 mW power, 4L-MO). The corneocytes from the stratum corneum can be identified. Supp info movie 1 shows a handheld holder where the endoscope distal head has been inserted to facilitate its manipulation and manual focusing for \textit{in vivo} imaging. 

\section{Imaging using the delivery fiber}

As mentioned above the stand-alone endoscope system can handle fs fiber laser systems at 920 nm, 780 nm and 1050 nm whose small footprints enable their integration into the endoscope kart. These $\sim$ 100 fs, 80 MHz laser repetition rate are suitable for 2-photon, SHG and possibly (1050 nm) THG imaging. However, there are situations where more energetic pulses are required such as for 3-photon and THG imaging. Typically 3-photon requires lower repetition rate $\sim$ 1 MHz and much shorter $\sim$ 50 fs  pulses at a typical wavelength of 1300 nm. Such energetic ($\sim$ $\mu$J) pulses can only be obtained from high-power oscillators pumping massive optical parametric amplifiers (OPAs). These laser systems are currently only available in a table-top format and cannot be integrated in a movable system.  In order to use our stand-alone cabinet for 3-photon and THG imaging, we developed a delivery fiber whose function is to deliver ultrashort pulses from a table-top laser system into the optical coupling unit of the endoscope kart (see Fig. \ref{fig:rack}(b)). The delivery fiber is a hollow core fiber similar to the one presented in Fig. \ref{fig:fibre}, except that it has a high-index polymer coating so that it does not have a double-clad. The transmission of the delivery fiber was measured to be 75\% across the 940 nm - 1300 nm range.   

\begin{figure}
    \centering
    \includegraphics[width=0.5\linewidth]{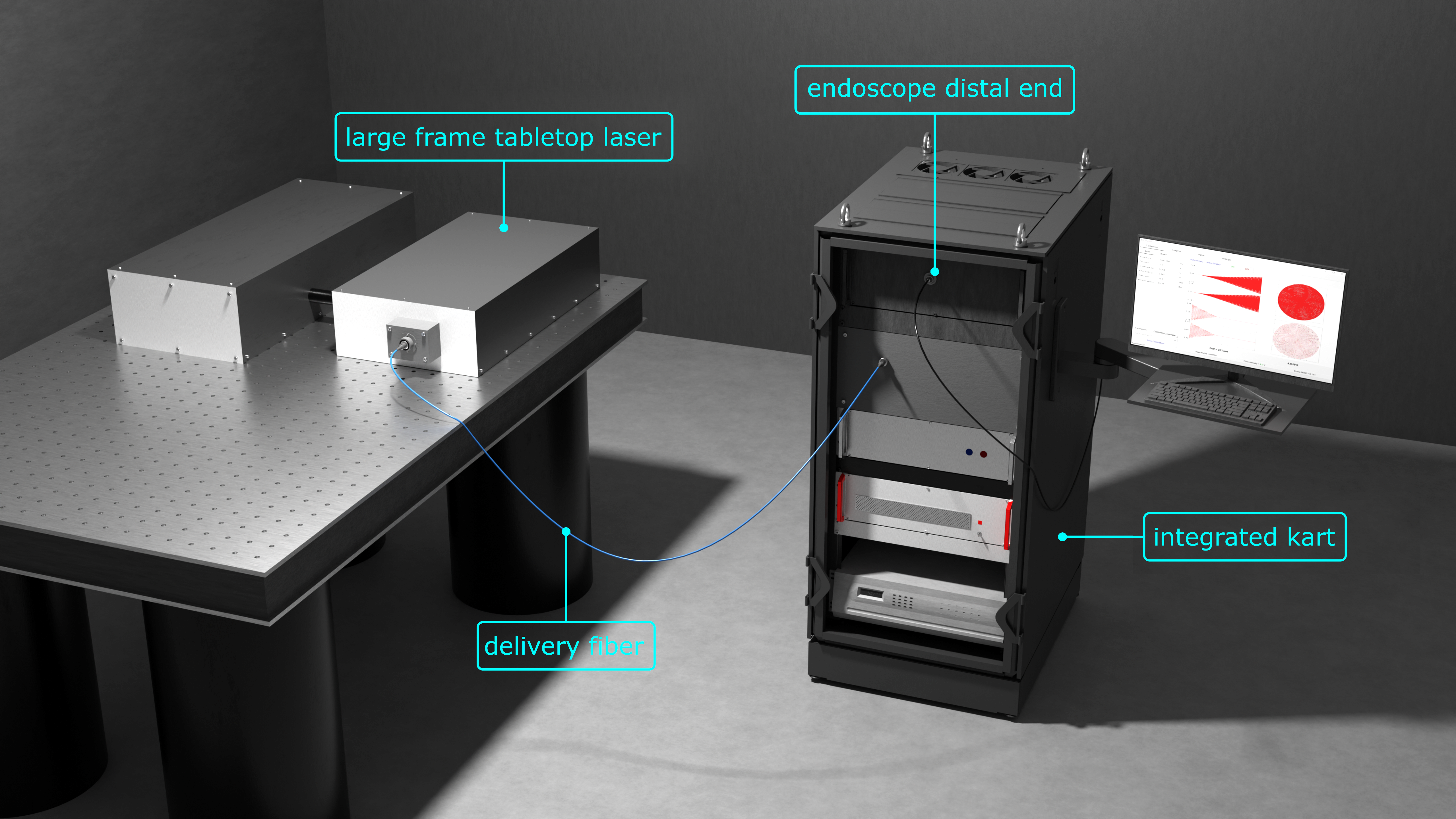}
    \caption{View of the delivery fiber scheme: a large frame table top laser light beam is brought into the endoscope kart system using the hollow core delivery fiber. }
    \label{fig:delivery}
\end{figure}

Fig. \ref{fig:delivery} shows a scheme of the experimental setup: a table-top laser system composed of a pump oscillator and an OPA is placed on an optical table. The laser beam is coupled into the delivery fiber whose output is plugged on the input port of the stand-alone kart. Then, the beam is coupled into the endoscopic probe using the optical coupling unit displayed in Fig. \ref{fig:rack}(b).  

\subsection{3-photon}
To perform 3-photon imaging we use an ytterbium-doped master oscillator power amplifier Monaco-1035 (Coherent) pumping and optical parametric amplifier (Opera-F, Coherent) that delivers 2.5 W of average power at 1300 nm and 940 nm with a 1 MHz repetition rate and $\sim$ 50 fs pulse duration. We use 100 mW of the OPA beam to get 75 mW at the delivery fiber output (fiber delivery transmission: 75\% ) that ended up to 30 mW at the endoscope distal side (endoscope transmission: 40\%). The pulse duration remained in the $\sim 50fs$ range at the sample plane.
\begin{figure}[!b]
    \centering
    \includegraphics[width=0.5\linewidth]{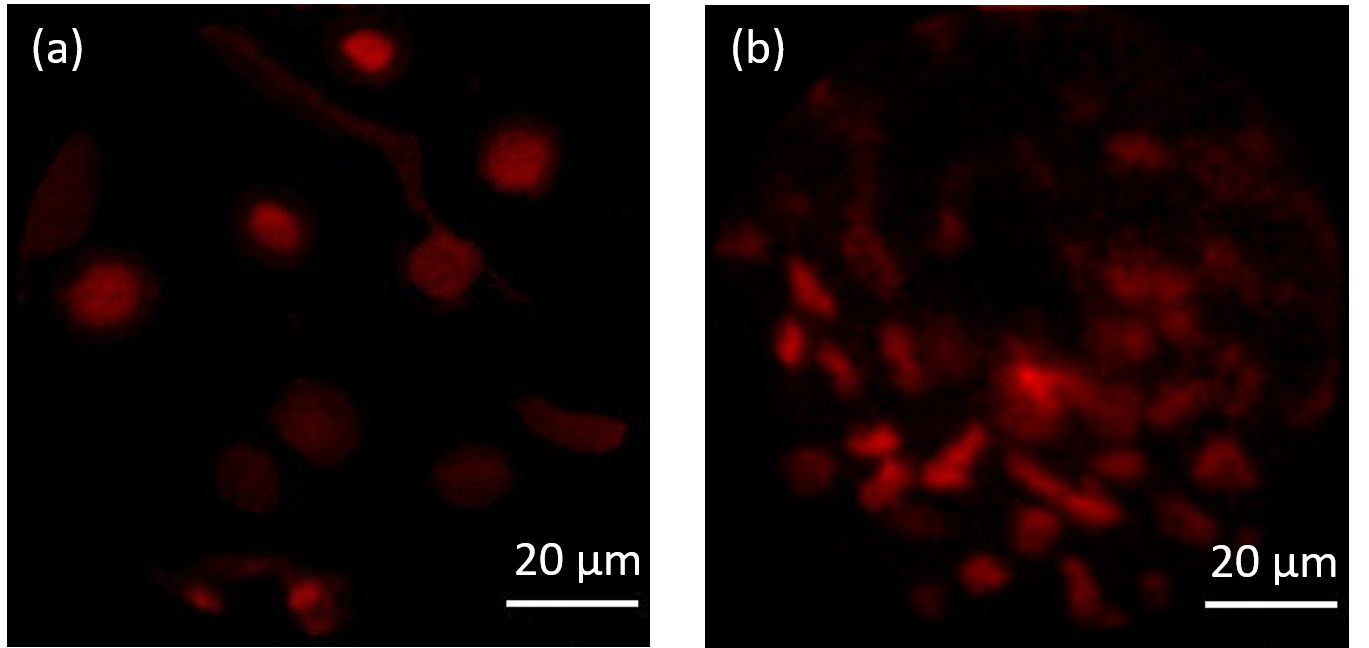}
    \caption{3-photon image of a DAPI stained pig retina (a) epithelium, (b) endothelium. See acquisition parameters in text. }
    \label{fig:3p}
\end{figure}
Figure \ref{fig:3p} (a) shows 3-photon fluorescence (excitation: 940 nm, detection: 433 nm) of a DAPI (4',6-diamidino-2-phenylindol) stained epithelium pig retina whereas Figure \ref{fig:3p} shows the endothelial of the same. Corneal button was carefully separated from the rest of the eyeball sample, stained with 10 µg/mL Hoechst 33342 solution in PBS for 5 min and rinsed in PBS without further fixation. Corneal button was directly placed in direct contact with a 35 mm glass bottom dish before observation. In both epithelium and endothelial parts the cell nuclei can be easily distinguished (FoV 98 µm, 1 fps, 30 mW average power, GRIN-MO).

\subsection{Multimodal imaging}
Using the built-in PMT detectors, the endoscope system can perform multi-color multimodal imaging.

\begin{figure}
    \centering
    \includegraphics[width=0.5\linewidth]{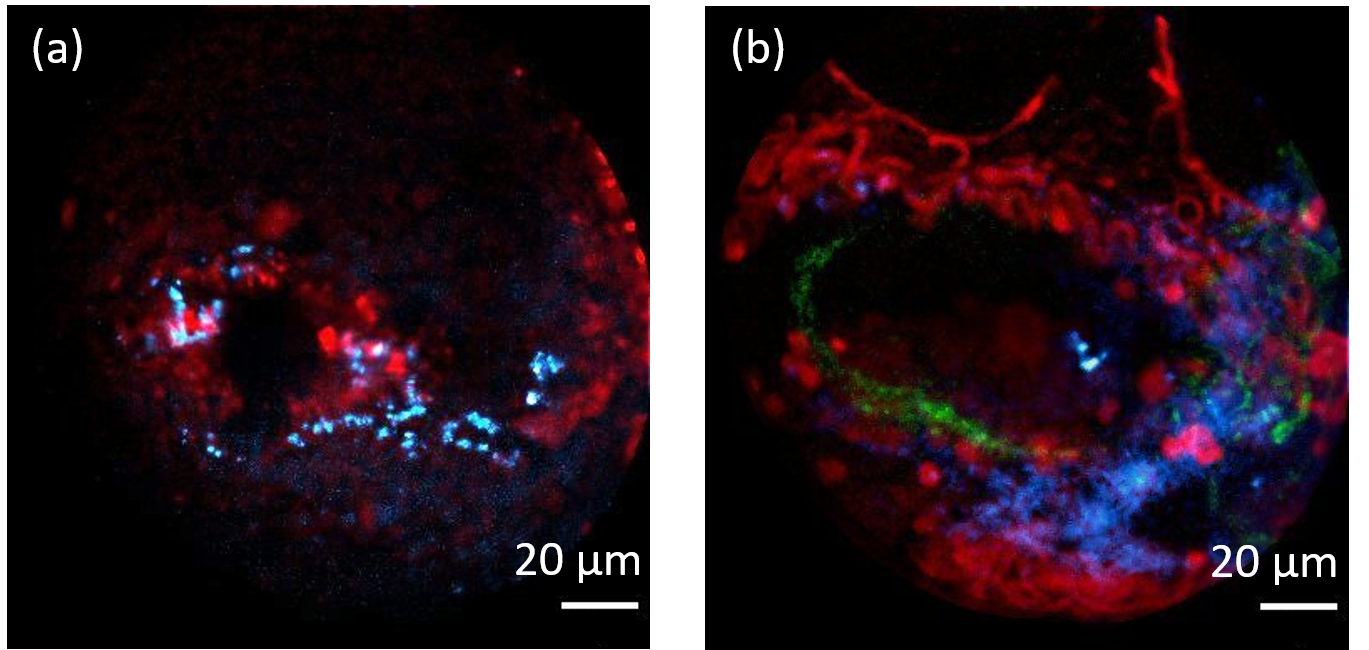}
    \caption{Multimodal imaging. (a) Composite 3-photon (red) and THG (blue) image of a human brain section stained with hematoxylin and eosin.  (b) 3-photon (red), SHG (green) and  THG (blue) image of a fixed de-fatted human adipocyte tissue stained with Masson's trichrome. See acquisition details in text. }
    \label{fig:multimodal}
\end{figure}

Figure \ref{fig:multimodal} (a) shows a 3-photon fluorescence (fluo: 520$\pm$30 nm) and  THG (detection: 433$\pm$20 nm) image of a human brain section stained with hematoxylin and eosin and excited at 1300 nm (FoV 168 µm, 1 fps, 15 mW power, GRIN-MO). The fluorescence comes from the eosin stain whereas the THG might come from nucleoli.  
Figure \ref{fig:multimodal} (b) shows a 3-photon fluorescence (red) (fluo: 520$\pm$30 nm), SHG (green) (detection: 650$\pm$30 nm) and  THG (blue) (detection: 433$\pm$20 nm)  image of a fixed de-fatted human adipocyte tissue stained with Masson's trichrome (Nublat commercial stained slides)  and excited at 1300 nm (FoV 168 µm, 1 frame/s, 15 mW power, GRIN-MO). Fluorescence reveals the adipocyte border whereas SHG reveals collagen fibers and THG small vesicles.

These images illustrate some of the imaging capabilities of the endoscope but some others could be added such as CARS \cite{lombardini_high-resolution_2018,septier_label-free_2022}.   

\section{Conclusion}
We have presented a flexible endoscope system based on a double-clad hollow core fiber that enables to perform nonlinear imaging over a broad range of near infrared excitation wavelenghs. We have demonstrated 2-photon/SHG imaging using a 920 nm, 100 fs laser and 3-photon/THG imaging using a 1300 nm, 50 fs high power laser system. Quite remarkably, the double-clad hollow core fiber can handle a broad range of excitation wavelength still maintaining the fs short pulse duration while the nonlinear signal is efficiently collected through the 0.38 NA double-cladding. The endoscope head as a distal diameter of 2.2 mm using the four lenses micro-objective (4L-MO) and 1.3 mm using the GRIN micro-objective (GRIN-MO) and can image across a FoV up to 600 µm at a maximum of 10 fps and with a $\sim$1 µm spatial resolution (depending on the micro-objective - see Table \ref{tab:table1}). The small endoscope head (weight $<$2 g, $\leq$2.2 mm diameter, 40 mm length) can be easily manipulated to access hard-to-reach places and we have demonstrated its handheld manipulation to image the skin forearm \textit{in vivo}. It could be also readily used for \textit{in vivo} brain activity monitoring in mice \cite{park_high-speed_2020}. The endoscope could be further improved by adding a $z$ scanner to the 2D resonant piezo scanner using small memory alloy actuators \cite{wu_fiber-optic_2010,li_focus_2017}. A unique feature brought by the weak nonlinearity of the hollow core fiber endoscope is its ability to activate nonlinear contrasts that involve overlapping pulses such as SRS \cite{lombardini_origin_2017} and CARS \cite{lombardini_high-resolution_2018}. We expect the endoscope to find a broad range of applications in the fields of biology, material sciences and medical applications as being a valuable miniature extension of bulky multiphoton microscopes.

\section*{Acknowledgments}

The fresh pig eyeball were graciously obtained from the euthanized pig (Département Hospitalo-Universitaire de Recherche et d’Enseignement - Dhure).\\
We acknowledge also financial support from the French National Research Agency (ANR-19-CE19-0019-03 MEAP, ANR-10-INSB-04-01, ANR-11-INSB-0006, ANR-16-CONV-0001, ANR-21-ESRS-0002 IDEC), the Contrats de Plan Etat-Region (CPER WaveTech), the French Ministry of Higher Education and Research, the Centre National de la Recherche Scientifique (CNRS), the Hauts-de-France (HdF) Regional Council, the European Regional Development Fund (ERDF), IRCICA.

{\appendix[Software]

Fig. \ref{fig:soft} shows some screenshots of the control endoscope software that has been developed to drive the stand-alone flexible endoscope system. In particular, it allows: to control the FoV, correct the dwell time issue, modulate the laser power to mitigate photobleaching, measure the signal in real-time, acquire and save images, perform image and video processing, color the images, adjustment the contrast, perform local motion correction, and fill blank pixels. It also performs the scanner calibration and allows to pre-record several calibrations. 
\begin{figure}[!h]
    \centering
    \includegraphics[width=0.5\linewidth]{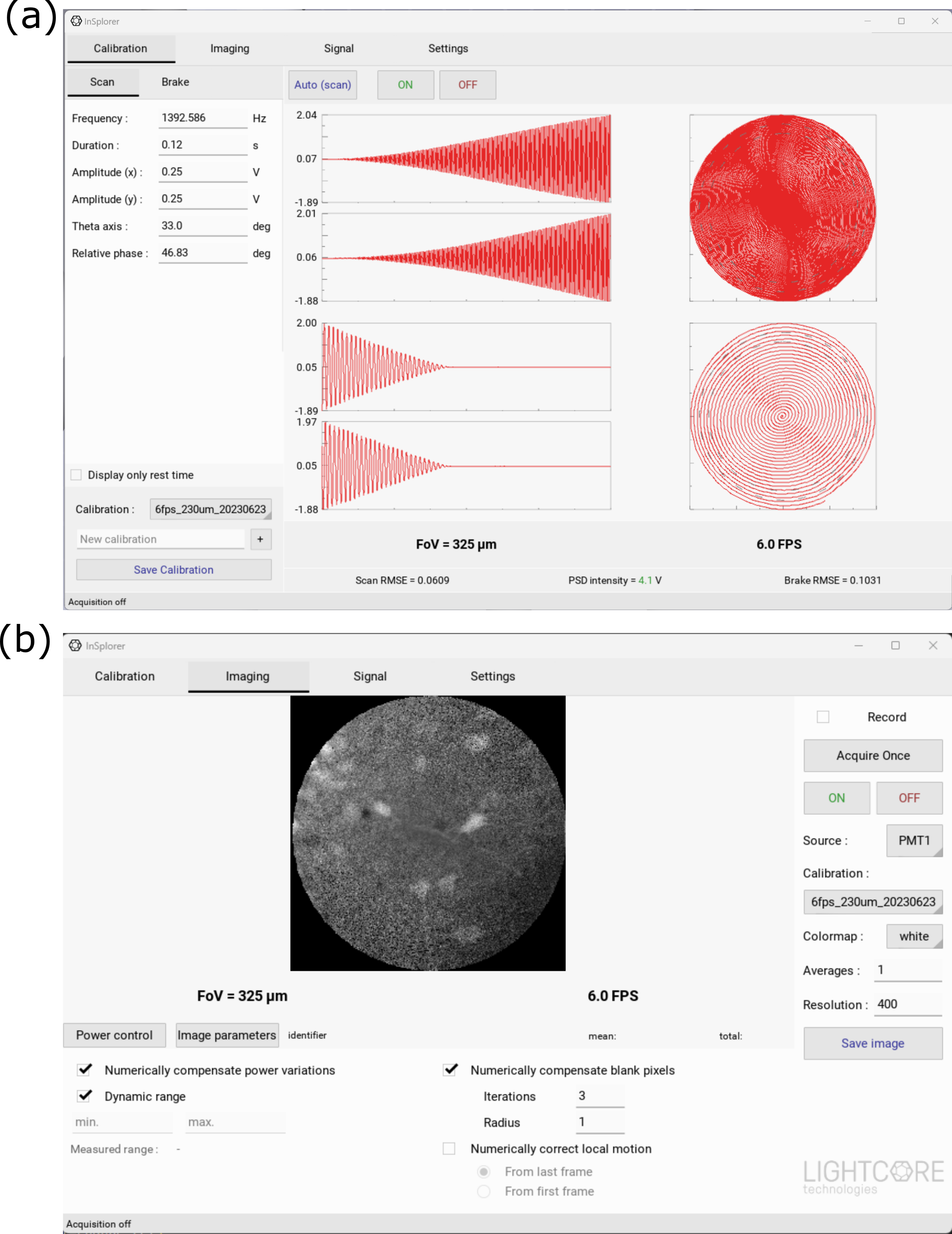}
    \caption{Screenshots of the custom control endoscope software showing the (a) 'calibration screen' with the scanner trajectories and the (b) 'imaging screen' with live-processing options.}
    \label{fig:soft}
\end{figure}

\bibliographystyle{IEEEtran}
\bibliography{JSTQE}

\begin{thebibliography}{10}
\providecommand{\url}[1]{#1}
\csname url@samestyle\endcsname
\providecommand{\newblock}{\relax}
\providecommand{\bibinfo}[2]{#2}
\providecommand{\BIBentrySTDinterwordspacing}{\spaceskip=0pt\relax}
\providecommand{\BIBentryALTinterwordstretchfactor}{4}
\providecommand{\BIBentryALTinterwordspacing}{\spaceskip=\fontdimen2\font plus
\BIBentryALTinterwordstretchfactor\fontdimen3\font minus \fontdimen4\font\relax}
\providecommand{\BIBforeignlanguage}[2]{{%
\expandafter\ifx\csname l@#1\endcsname\relax
\typeout{** WARNING: IEEEtran.bst: No hyphenation pattern has been}%
\typeout{** loaded for the language `#1'. Using the pattern for}%
\typeout{** the default language instead.}%
\else
\language=\csname l@#1\endcsname
\fi
#2}}
\providecommand{\BIBdecl}{\relax}
\BIBdecl

\bibitem{zipfel_live_2003}
\BIBentryALTinterwordspacing
W.~R. Zipfel, R.~M. Williams, R.~Christie, A.~Y. Nikitin, B.~T. Hyman, and W.~W. Webb, ``Live {Tissue} {Intrinsic} {Emission} {Microscopy} {Using} {Multiphoton}-{Excited} {Native} {Fluorescence} and {Second} {Harmonic} {Generation},'' \emph{Proceedings of the National Academy of Sciences of the United States of America}, vol. 100, no.~12, pp. 7075--7080, 2003, publisher: National Academy of Sciences. [Online]. Available: \url{https://www.jstor.org/stable/3139434}
\BIBentrySTDinterwordspacing

\bibitem{sheppard_multiphoton_2020}
\BIBentryALTinterwordspacing
C.~J.~R. Sheppard, ``Multiphoton microscopy: a personal historical review, with some future predictions,'' \emph{Journal of Biomedical Optics}, vol.~25, no.~1, p. 014511, Jan. 2020, publisher: SPIE. [Online]. Available: \url{https://www.spiedigitallibrary.org/journals/journal-of-biomedical-optics/volume-25/issue-1/014511/Multiphoton-microscopy--a-personal-historical-review-with-some-future/10.1117/1.JBO.25.1.014511.full}
\BIBentrySTDinterwordspacing

\bibitem{carriles_invited_2009}
\BIBentryALTinterwordspacing
R.~Carriles, D.~N. Schafer, K.~E. Sheetz, J.~J. Field, R.~Cisek, V.~Barzda, A.~W. Sylvester, and J.~A. Squier, ``Invited {Review} {Article}: {Imaging} techniques for harmonic and multiphoton absorption fluorescence microscopy,'' \emph{Review of Scientific Instruments}, vol.~80, no.~8, p. 081101, Aug. 2009. [Online]. Available: \url{https://doi.org/10.1063/1.3184828}
\BIBentrySTDinterwordspacing

\bibitem{miller_deep_2017}
\BIBentryALTinterwordspacing
D.~R. Miller, J.~W. Jarrett, A.~M. Hassan, and A.~K. Dunn, ``Deep tissue imaging with multiphoton fluorescence microscopy,'' \emph{Current Opinion in Biomedical Engineering}, vol.~4, pp. 32--39, Dec. 2017. [Online]. Available: \url{https://www.sciencedirect.com/science/article/pii/S2468451117300466}
\BIBentrySTDinterwordspacing

\bibitem{diaspro_multi-photon_2006}
\BIBentryALTinterwordspacing
A.~Diaspro, P.~Bianchini, G.~Vicidomini, M.~Faretta, P.~Ramoino, and C.~Usai, ``Multi-photon excitation microscopy,'' \emph{BioMedical Engineering OnLine}, vol.~5, no.~1, p.~36, Jun. 2006. [Online]. Available: \url{https://doi.org/10.1186/1475-925X-5-36}
\BIBentrySTDinterwordspacing

\bibitem{ranawat_recent_2019}
H.~Ranawat, S.~Pal, and N.~Mazumder, ``\BIBforeignlanguage{eng}{Recent trends in two-photon auto-fluorescence lifetime imaging ({2P}-{FLIM}) and its biomedical applications},'' \emph{\BIBforeignlanguage{eng}{Biomedical Engineering Letters}}, vol.~9, no.~3, pp. 293--310, Aug. 2019.

\bibitem{aghigh_second_2023}
A.~Aghigh, S.~Bancelin, M.~Rivard, M.~Pinsard, H.~Ibrahim, and F.~Légaré, ``\BIBforeignlanguage{eng}{Second harmonic generation microscopy: a powerful tool for bio-imaging},'' \emph{\BIBforeignlanguage{eng}{Biophysical Reviews}}, vol.~15, no.~1, pp. 43--70, Feb. 2023.

\bibitem{weigelin_third_2016}
B.~Weigelin, G.-J. Bakker, and P.~Friedl, ``\BIBforeignlanguage{eng}{Third harmonic generation microscopy of cells and tissue organization},'' \emph{\BIBforeignlanguage{eng}{Journal of Cell Science}}, vol. 129, no.~2, pp. 245--255, Jan. 2016.

\bibitem{min_near-degenerate_2009}
\BIBentryALTinterwordspacing
W.~Min, S.~Lu, M.~Rueckel, G.~R. Holtom, and X.~S. Xie, ``Near-{Degenerate} {Four}-{Wave}-{Mixing} {Microscopy},'' \emph{Nano Letters}, vol.~9, no.~6, pp. 2423--2426, Jun. 2009, publisher: American Chemical Society. [Online]. Available: \url{https://doi.org/10.1021/nl901101g}
\BIBentrySTDinterwordspacing

\bibitem{min_coherent_2011}
\BIBentryALTinterwordspacing
W.~Min, C.~W. Freudiger, S.~Lu, and X.~S. Xie, ``Coherent {Nonlinear} {Optical} {Imaging}: {Beyond} {Fluorescence} {Microscopy},'' \emph{Annual Review of Physical Chemistry}, vol.~62, no.~1, pp. 507--530, 2011, \_eprint: https://doi.org/10.1146/annurev.physchem.012809.103512. [Online]. Available: \url{https://doi.org/10.1146/annurev.physchem.012809.103512}
\BIBentrySTDinterwordspacing

\bibitem{rigneault_tutorial_2018}
\BIBentryALTinterwordspacing
H.~Rigneault and P.~Berto, ``Tutorial: {Coherent} {Raman} light matter interaction processes,'' \emph{APL Photonics}, vol.~3, no.~9, p. 091101, Jul. 2018. [Online]. Available: \url{https://doi.org/10.1063/1.5030335}
\BIBentrySTDinterwordspacing

\bibitem{fischer_invited_2016}
\BIBentryALTinterwordspacing
M.~C. Fischer, J.~W. Wilson, F.~E. Robles, and W.~S. Warren, ``Invited {Review} {Article}: {Pump}-probe microscopy,'' \emph{Review of Scientific Instruments}, vol.~87, no.~3, p. 031101, Mar. 2016. [Online]. Available: \url{https://doi.org/10.1063/1.4943211}
\BIBentrySTDinterwordspacing

\bibitem{ishii_focusing_2022}
\BIBentryALTinterwordspacing
H.~Ishii, K.~Otomo, T.~Takahashi, K.~Yamaguchi, and T.~Nemoto, ``Focusing new light on brain functions: multiphoton microscopy for deep and super-resolution imaging,'' \emph{Neuroscience Research}, vol. 179, pp. 24--30, Jun. 2022. [Online]. Available: \url{https://www.sciencedirect.com/science/article/pii/S0168010221002455}
\BIBentrySTDinterwordspacing

\bibitem{konig_multiphoton_2000}
\BIBentryALTinterwordspacing
K.~König, ``\BIBforeignlanguage{en}{Multiphoton microscopy in life sciences},'' \emph{\BIBforeignlanguage{en}{Journal of Microscopy}}, vol. 200, no.~2, pp. 83--104, 2000, \_eprint: https://onlinelibrary.wiley.com/doi/pdf/10.1046/j.1365-2818.2000.00738.x. [Online]. Available: \url{https://onlinelibrary.wiley.com/doi/abs/10.1046/j.1365-2818.2000.00738.x}
\BIBentrySTDinterwordspacing

\bibitem{wang_two-photon_2010}
\BIBentryALTinterwordspacing
B.-G. Wang, K.~König, and K.-J. Halbhuber, ``\BIBforeignlanguage{en}{Two-photon microscopy of deep intravital tissues and its merits in clinical research},'' \emph{\BIBforeignlanguage{en}{Journal of Microscopy}}, vol. 238, no.~1, pp. 1--20, 2010, \_eprint: https://onlinelibrary.wiley.com/doi/pdf/10.1111/j.1365-2818.2009.03330.x. [Online]. Available: \url{https://onlinelibrary.wiley.com/doi/abs/10.1111/j.1365-2818.2009.03330.x}
\BIBentrySTDinterwordspacing

\bibitem{orringer_rapid_2017}
\BIBentryALTinterwordspacing
D.~A. Orringer, B.~Pandian, Y.~S. Niknafs, T.~C. Hollon, J.~Boyle, S.~Lewis, M.~Garrard, S.~L. Hervey-Jumper, H.~J.~L. Garton, C.~O. Maher, J.~A. Heth, O.~Sagher, D.~A. Wilkinson, M.~Snuderl, S.~Venneti, S.~H. Ramkissoon, K.~A. McFadden, A.~Fisher-Hubbard, A.~P. Lieberman, T.~D. Johnson, X.~S. Xie, J.~K. Trautman, C.~W. Freudiger, and S.~Camelo-Piragua, ``\BIBforeignlanguage{en}{Rapid intraoperative histology of unprocessed surgical specimens via fibre-laser-based stimulated {Raman} scattering microscopy},'' \emph{\BIBforeignlanguage{en}{Nature Biomedical Engineering}}, vol.~1, no.~2, pp. 1--13, Feb. 2017, number: 2 Publisher: Nature Publishing Group. [Online]. Available: \url{https://www.nature.com/articles/s41551-016-0027}
\BIBentrySTDinterwordspacing

\bibitem{yue_multimodal_2011}
\BIBentryALTinterwordspacing
S.~Yue, M.~Slipchenko, and J.-X. Cheng, ``\BIBforeignlanguage{en}{Multimodal nonlinear optical microscopy},'' \emph{\BIBforeignlanguage{en}{Laser \& Photonics Reviews}}, vol.~5, no.~4, pp. 496--512, 2011, \_eprint: https://onlinelibrary.wiley.com/doi/pdf/10.1002/lpor.201000027. [Online]. Available: \url{https://onlinelibrary.wiley.com/doi/abs/10.1002/lpor.201000027}
\BIBentrySTDinterwordspacing

\bibitem{tang_design_2009}
\BIBentryALTinterwordspacing
S.~Tang, W.~Jung, D.~McCormick, T.~Xie, J.~Su, Y.-C. Ahn, B.~J. Tromberg, and Z.~Chen, ``Design and implementation of fiber-based multiphoton endoscopy with microelectromechanical systems scanning,'' \emph{Journal of biomedical optics}, vol.~14, no.~3, p. 034005, 2009. [Online]. Available: \url{https://www.ncbi.nlm.nih.gov/pmc/articles/PMC2866630/}
\BIBentrySTDinterwordspacing

\bibitem{myaing_fiber-optic_2006}
\BIBentryALTinterwordspacing
M.~T. Myaing, D.~J. MacDonald, and X.~Li, ``\BIBforeignlanguage{EN}{Fiber-optic scanning two-photon fluorescence endoscope},'' \emph{\BIBforeignlanguage{EN}{Optics Letters}}, vol.~31, no.~8, pp. 1076--1078, Apr. 2006, publisher: Optica Publishing Group. [Online]. Available: \url{https://opg.optica.org/ol/abstract.cfm?uri=ol-31-8-1076}
\BIBentrySTDinterwordspacing

\bibitem{do_fiber-optic_2014}
D.~Do, H.~Yoo, and D.-G. Gweon, ``\BIBforeignlanguage{eng}{Fiber-optic raster scanning two-photon endomicroscope using a tubular piezoelectric actuator},'' \emph{\BIBforeignlanguage{eng}{Journal of Biomedical Optics}}, vol.~19, no.~6, p. 066010, Jun. 2014.

\bibitem{rivera_use_2012}
\BIBentryALTinterwordspacing
D.~R. Rivera, C.~M. Brown, D.~G. Ouzounov, W.~W. Webb, and C.~Xu, ``\BIBforeignlanguage{EN}{Use of a lensed fiber for a large-field-of-view, high-resolution, fiber-scanning microendoscope},'' \emph{\BIBforeignlanguage{EN}{Optics Letters}}, vol.~37, no.~5, pp. 881--883, Mar. 2012, publisher: Optica Publishing Group. [Online]. Available: \url{https://opg.optica.org/ol/abstract.cfm?uri=ol-37-5-881}
\BIBentrySTDinterwordspacing

\bibitem{lukic_endoscopic_2017}
\BIBentryALTinterwordspacing
A.~Lukic, S.~Dochow, H.~Bae, G.~Matz, I.~Latka, B.~Messerschmidt, M.~Schmitt, and J.~Popp, ``\BIBforeignlanguage{EN}{Endoscopic fiber probe for nonlinear spectroscopic imaging},'' \emph{\BIBforeignlanguage{EN}{Optica}}, vol.~4, no.~5, pp. 496--501, May 2017, publisher: Optica Publishing Group. [Online]. Available: \url{https://opg.optica.org/optica/abstract.cfm?uri=optica-4-5-496}
\BIBentrySTDinterwordspacing

\bibitem{wu_scanning_2009}
\BIBentryALTinterwordspacing
Y.~Wu, Y.~Leng, J.~Xi, and X.~Li, ``\BIBforeignlanguage{EN}{Scanning all-fiber-optic endomicroscopy system for {3D} nonlinear optical imaging of biological tissues},'' \emph{\BIBforeignlanguage{EN}{Optics Express}}, vol.~17, no.~10, pp. 7907--7915, May 2009, publisher: Optica Publishing Group. [Online]. Available: \url{https://opg.optica.org/oe/abstract.cfm?uri=oe-17-10-7907}
\BIBentrySTDinterwordspacing

\bibitem{murari_compensation-free_2011}
\BIBentryALTinterwordspacing
K.~Murari, Y.~Zhang, S.~Li, Y.~Chen, M.-J. Li, and X.~Li, ``\BIBforeignlanguage{EN}{Compensation-free, all-fiber-optic, two-photon endomicroscopy at 1.55 $\mu$m},'' \emph{\BIBforeignlanguage{EN}{Optics Letters}}, vol.~36, no.~7, pp. 1299--1301, Apr. 2011, publisher: Optica Publishing Group. [Online]. Available: \url{https://opg.optica.org/ol/abstract.cfm?uri=ol-36-7-1299}
\BIBentrySTDinterwordspacing

\bibitem{ducourthial_development_2015}
\BIBentryALTinterwordspacing
G.~Ducourthial, P.~Leclerc, T.~Mansuryan, M.~Fabert, J.~Brevier, R.~Habert, F.~Braud, R.~Batrin, C.~Vever-Bizet, G.~Bourg-Heckly, L.~Thiberville, A.~Druilhe, A.~Kudlinski, and F.~Louradour, ``\BIBforeignlanguage{en}{Development of a real-time flexible multiphoton microendoscope for label-free imaging in a live animal},'' \emph{\BIBforeignlanguage{en}{Scientific Reports}}, vol.~5, no.~1, p. 18303, Dec. 2015, number: 1 Publisher: Nature Publishing Group. [Online]. Available: \url{https://www.nature.com/articles/srep18303}
\BIBentrySTDinterwordspacing

\bibitem{lombardini_high-resolution_2018}
\BIBentryALTinterwordspacing
A.~Lombardini, V.~Mytskaniuk, S.~Sivankutty, E.~R. Andresen, X.~Chen, J.~Wenger, M.~Fabert, N.~Joly, F.~Louradour, A.~Kudlinski, and H.~Rigneault, ``\BIBforeignlanguage{en}{High-resolution multimodal flexible coherent {Raman} endoscope},'' \emph{\BIBforeignlanguage{en}{Light: Science \& Applications}}, vol.~7, no.~1, p.~10, May 2018, number: 1 Publisher: Nature Publishing Group. [Online]. Available: \url{https://www.nature.com/articles/s41377-018-0003-3}
\BIBentrySTDinterwordspacing

\bibitem{kudlinski_double_2020}
\BIBentryALTinterwordspacing
A.~Kudlinski, A.~Cassez, O.~Vanvincq, D.~Septier, A.~Pastre, R.~Habert, K.~Baudelle, M.~Douay, V.~Mytskaniuk, V.~Tsvirkun, H.~Rigneault, and G.~Bouwmans, ``\BIBforeignlanguage{EN}{Double clad tubular anti-resonant hollow core fiber for nonlinear microendoscopy},'' \emph{\BIBforeignlanguage{EN}{Optics Express}}, vol.~28, no.~10, pp. 15\,062--15\,070, May 2020, publisher: Optica Publishing Group. [Online]. Available: \url{https://opg.optica.org/oe/abstract.cfm?uri=oe-28-10-15062}
\BIBentrySTDinterwordspacing

\bibitem{pshenay-severin_multimodal_2021}
\BIBentryALTinterwordspacing
E.~Pshenay-Severin, H.~Bae, K.~Reichwald, G.~Matz, J.~Bierlich, J.~Kobelke, A.~Lorenz, A.~Schwuchow, T.~Meyer-Zedler, M.~Schmitt, B.~Messerschmidt, and J.~Popp, ``\BIBforeignlanguage{en}{Multimodal nonlinear endomicroscopic imaging probe using a double-core double-clad fiber and focus-combining micro-optical concept},'' \emph{\BIBforeignlanguage{en}{Light: Science \& Applications}}, vol.~10, no.~1, p. 207, Oct. 2021, number: 1 Publisher: Nature Publishing Group. [Online]. Available: \url{https://www.nature.com/articles/s41377-021-00648-w}
\BIBentrySTDinterwordspacing

\bibitem{septier_label-free_2022}
\BIBentryALTinterwordspacing
D.~Septier, V.~Mytskaniuk, R.~Habert, D.~Labat, K.~Baudelle, A.~Cassez, G.~Brévalle-Wasilewski, M.~Conforti, G.~Bouwmans, H.~Rigneault, and A.~Kudlinski, ``\BIBforeignlanguage{EN}{Label-free highly multimodal nonlinear endoscope},'' \emph{\BIBforeignlanguage{EN}{Optics Express}}, vol.~30, no.~14, pp. 25\,020--25\,033, Jul. 2022, publisher: Optica Publishing Group. [Online]. Available: \url{https://opg.optica.org/oe/abstract.cfm?uri=oe-30-14-25020}
\BIBentrySTDinterwordspacing

\bibitem{kim_lissajous_2019}
\BIBentryALTinterwordspacing
D.~Y. Kim, K.~Hwang, J.~Ahn, Y.-H. Seo, J.-B. Kim, S.~Lee, J.-H. Yoon, E.~Kong, Y.~Jeong, S.~Jon, P.~Kim, and K.-H. Jeong, ``\BIBforeignlanguage{en}{Lissajous {Scanning} {Two}-photon {Endomicroscope} for {In} vivo {Tissue} {Imaging}},'' \emph{\BIBforeignlanguage{en}{Scientific Reports}}, vol.~9, no.~1, p. 3560, Mar. 2019, number: 1 Publisher: Nature Publishing Group. [Online]. Available: \url{https://www.nature.com/articles/s41598-019-38762-w}
\BIBentrySTDinterwordspacing

\bibitem{zhang_compact_2012}
\BIBentryALTinterwordspacing
Y.~Zhang, M.~L. Akins, K.~Murari, J.~Xi, M.-J. Li, K.~Luby-Phelps, M.~Mahendroo, and X.~Li, ``A compact fiber-optic {SHG} scanning endomicroscope and its application to visualize cervical remodeling during pregnancy,'' \emph{Proceedings of the National Academy of Sciences}, vol. 109, no.~32, pp. 12\,878--12\,883, Aug. 2012, publisher: Proceedings of the National Academy of Sciences. [Online]. Available: \url{https://www.pnas.org/doi/abs/10.1073/pnas.1121495109}
\BIBentrySTDinterwordspacing

\bibitem{zhao_development_2010}
\BIBentryALTinterwordspacing
Y.~Zhao, H.~Nakamura, and R.~J. Gordon, ``\BIBforeignlanguage{EN}{Development of a versatile two-photon endoscope for biological imaging},'' \emph{\BIBforeignlanguage{EN}{Biomedical Optics Express}}, vol.~1, no.~4, pp. 1159--1172, Nov. 2010, publisher: Optica Publishing Group. [Online]. Available: \url{https://opg.optica.org/boe/abstract.cfm?uri=boe-1-4-1159}
\BIBentrySTDinterwordspacing

\bibitem{akhoundi_compact_2018}
\BIBentryALTinterwordspacing
F.~Akhoundi, Y.~Qin, N.~Peyghambarian, J.~K. Barton, and K.~Kieu, ``\BIBforeignlanguage{EN}{Compact fiber-based multi-photon endoscope working at 1700 nm},'' \emph{\BIBforeignlanguage{EN}{Biomedical Optics Express}}, vol.~9, no.~5, pp. 2326--2335, May 2018, publisher: Optica Publishing Group. [Online]. Available: \url{https://opg.optica.org/boe/abstract.cfm?uri=boe-9-5-2326}
\BIBentrySTDinterwordspacing

\bibitem{fu_nonlinear_2006}
\BIBentryALTinterwordspacing
L.~Fu, A.~Jain, H.~Xie, C.~Cranfield, and M.~Gu, ``\BIBforeignlanguage{EN}{Nonlinear optical endoscopy based on a double-clad photonic crystal fiber and a {MEMS} mirror},'' \emph{\BIBforeignlanguage{EN}{Optics Express}}, vol.~14, no.~3, pp. 1027--1032, Feb. 2006, publisher: Optica Publishing Group. [Online]. Available: \url{https://opg.optica.org/oe/abstract.cfm?uri=oe-14-3-1027}
\BIBentrySTDinterwordspacing

\bibitem{chang_two-photon_2008}
\BIBentryALTinterwordspacing
Y.-C. Chang, J.~Y. Ye, T.~Thomas, Y.~Chen, J.~R. Baker, and T.~B. Norris, ``\BIBforeignlanguage{EN}{Two-photon fluorescence correlation spectroscopy through a dual-clad optical fiber},'' \emph{\BIBforeignlanguage{EN}{Optics Express}}, vol.~16, no.~17, pp. 12\,640--12\,649, Aug. 2008, publisher: Optica Publishing Group. [Online]. Available: \url{https://opg.optica.org/oe/abstract.cfm?uri=oe-16-17-12640}
\BIBentrySTDinterwordspacing

\bibitem{rivera_compact_2011}
\BIBentryALTinterwordspacing
D.~R. Rivera, C.~M. Brown, D.~G. Ouzounov, I.~Pavlova, D.~Kobat, W.~W. Webb, and C.~Xu, ``Compact and flexible raster scanning multiphoton endoscope capable of imaging unstained tissue,'' \emph{Proceedings of the National Academy of Sciences}, vol. 108, no.~43, pp. 17\,598--17\,603, Oct. 2011, publisher: Proceedings of the National Academy of Sciences. [Online]. Available: \url{https://www.pnas.org/doi/full/10.1073/pnas.1114746108}
\BIBentrySTDinterwordspacing

\bibitem{li_twist-free_2021}
\BIBentryALTinterwordspacing
A.~Li, H.~Guan, H.-C. Park, Y.~Yue, D.~Chen, W.~Liang, M.-J. Li, H.~Lu, and X.~Li, ``\BIBforeignlanguage{EN}{Twist-free ultralight two-photon fiberscope enabling neuroimaging on freely rotating/walking mice},'' \emph{\BIBforeignlanguage{EN}{Optica}}, vol.~8, no.~6, pp. 870--879, Jun. 2021, publisher: Optica Publishing Group. [Online]. Available: \url{https://opg.optica.org/optica/abstract.cfm?uri=optica-8-6-870}
\BIBentrySTDinterwordspacing

\bibitem{kucikas_two-photon_2023}
\BIBentryALTinterwordspacing
V.~Kučikas, M.~P. Werner, T.~Schmitz-Rode, F.~Louradour, and M.~A. M.~J. van Zandvoort, ``\BIBforeignlanguage{en}{Two-{Photon} {Endoscopy}: {State} of the {Art} and {Perspectives}},'' \emph{\BIBforeignlanguage{en}{Molecular Imaging and Biology}}, vol.~25, no.~1, pp. 3--17, Feb. 2023. [Online]. Available: \url{https://doi.org/10.1007/s11307-021-01665-2}
\BIBentrySTDinterwordspacing

\bibitem{klioutchnikov_three-photon_2020}
\BIBentryALTinterwordspacing
A.~Klioutchnikov, D.~J. Wallace, M.~H. Frosz, R.~Zeltner, J.~Sawinski, V.~Pawlak, K.-M. Voit, P.~S.~J. Russell, and J.~N.~D. Kerr, ``\BIBforeignlanguage{en}{Three-photon head-mounted microscope for imaging deep cortical layers in freely moving rats},'' \emph{\BIBforeignlanguage{en}{Nature Methods}}, vol.~17, no.~5, pp. 509--513, May 2020, number: 5 Publisher: Nature Publishing Group. [Online]. Available: \url{https://www.nature.com/articles/s41592-020-0817-9}
\BIBentrySTDinterwordspacing

\bibitem{huland_three-photon_2013}
\BIBentryALTinterwordspacing
D.~M. Huland, K.~Charan, D.~G. Ouzounov, J.~S. Jones, N.~Nishimura, and C.~Xu, ``\BIBforeignlanguage{EN}{Three-photon excited fluorescence imaging of unstained tissue using a {GRIN} lens endoscope},'' \emph{\BIBforeignlanguage{EN}{Biomedical Optics Express}}, vol.~4, no.~5, pp. 652--658, May 2013, publisher: Optica Publishing Group. [Online]. Available: \url{https://opg.optica.org/boe/abstract.cfm?uri=boe-4-5-652}
\BIBentrySTDinterwordspacing

\bibitem{kolyadin_negative_2015}
\BIBentryALTinterwordspacing
A.~N. Kolyadin, G.~K. Alagashev, A.~D. Pryamikov, L.~Mouradian, A.~Zeytunyan, H.~Toneyan, A.~F. Kosolapov, and I.~A. Bufetov, ``Negative {Curvature} {Hollow}-core {Fibers}: {Dispersion} {Properties} and {Femtosecond} {Pulse} {Delivery},'' \emph{Physics Procedia}, vol.~73, pp. 59--66, Jan. 2015. [Online]. Available: \url{https://www.sciencedirect.com/science/article/pii/S1875389215012924}
\BIBentrySTDinterwordspacing

\bibitem{wei_negative_2017}
\BIBentryALTinterwordspacing
C.~Wei, R.~J. Weiblen, C.~R. Menyuk, and J.~Hu, ``\BIBforeignlanguage{EN}{Negative curvature fibers},'' \emph{\BIBforeignlanguage{EN}{Advances in Optics and Photonics}}, vol.~9, no.~3, pp. 504--561, Sep. 2017, publisher: Optica Publishing Group. [Online]. Available: \url{https://opg.optica.org/aop/abstract.cfm?uri=aop-9-3-504}
\BIBentrySTDinterwordspacing

\bibitem{debord_ultralow_2017}
\BIBentryALTinterwordspacing
B.~Debord, A.~Amsanpally, M.~Chafer, A.~Baz, M.~Maurel, J.~M. Blondy, E.~Hugonnot, F.~Scol, L.~Vincetti, F.~Gérôme, and F.~Benabid, ``\BIBforeignlanguage{EN}{Ultralow transmission loss in inhibited-coupling guiding hollow fibers},'' \emph{\BIBforeignlanguage{EN}{Optica}}, vol.~4, no.~2, pp. 209--217, Feb. 2017, publisher: Optica Publishing Group. [Online]. Available: \url{https://opg.optica.org/optica/abstract.cfm?uri=optica-4-2-209}
\BIBentrySTDinterwordspacing

\bibitem{tateda_interferometric_1981}
\BIBentryALTinterwordspacing
M.~Tateda, N.~Shibata, and S.~Seikai, ``Interferometric method for chromatic dispersion measurement in a single-mode optical fiber,'' \emph{IEEE Journal of Quantum Electronics}, vol.~17, no.~3, pp. 404--407, Mar. 1981, conference Name: IEEE Journal of Quantum Electronics. [Online]. Available: \url{https://ieeexplore.ieee.org/abstract/document/1071115}
\BIBentrySTDinterwordspacing

\bibitem{zeisberger_analytic_2017}
\BIBentryALTinterwordspacing
M.~Zeisberger and M.~A. Schmidt, ``\BIBforeignlanguage{en}{Analytic model for the complex effective index of the leaky modes of tube-type anti-resonant hollow core fibers},'' \emph{\BIBforeignlanguage{en}{Scientific Reports}}, vol.~7, no.~1, p. 11761, Sep. 2017, number: 1 Publisher: Nature Publishing Group. [Online]. Available: \url{https://www.nature.com/articles/s41598-017-12234-5}
\BIBentrySTDinterwordspacing

\bibitem{iga_theory_1980}
\BIBentryALTinterwordspacing
K.~Iga, ``\BIBforeignlanguage{EN}{Theory for gradient-index imaging},'' \emph{\BIBforeignlanguage{EN}{Applied Optics}}, vol.~19, no.~7, pp. 1039--1043, Apr. 1980, publisher: Optica Publishing Group. [Online]. Available: \url{https://opg.optica.org/ao/abstract.cfm?uri=ao-19-7-1039}
\BIBentrySTDinterwordspacing

\bibitem{thirion_image_1998}
\BIBentryALTinterwordspacing
J.~P. Thirion, ``Image matching as a diffusion process: an analogy with {Maxwell}'s demons,'' \emph{Medical Image Analysis}, vol.~2, no.~3, pp. 243--260, Sep. 1998. [Online]. Available: \url{https://www.sciencedirect.com/science/article/pii/S1361841598800224}
\BIBentrySTDinterwordspacing

\bibitem{park_high-speed_2020}
\BIBentryALTinterwordspacing
H.-C. Park, H.~Guan, A.~Li, Y.~Yue, M.-J. Li, H.~Lu, and X.~Li, ``\BIBforeignlanguage{EN}{High-speed fiber-optic scanning nonlinear endomicroscopy for imaging neuron dynamics in vivo},'' \emph{\BIBforeignlanguage{EN}{Optics Letters}}, vol.~45, no.~13, pp. 3605--3608, Jul. 2020, publisher: Optica Publishing Group. [Online]. Available: \url{https://opg.optica.org/ol/abstract.cfm?uri=ol-45-13-3605}
\BIBentrySTDinterwordspacing

\bibitem{wu_fiber-optic_2010}
\BIBentryALTinterwordspacing
Y.~Wu, Y.~Zhang, J.~Xi, M.-J. Li, and X.~Li, ``Fiber-optic nonlinear endomicroscopy with focus scanning by using shape memory alloy actuation,'' \emph{Journal of Biomedical Optics}, vol.~15, no.~6, p. 060506, Nov. 2010, publisher: SPIE. [Online]. Available: \url{https://www.spiedigitallibrary.org/journals/journal-of-biomedical-optics/volume-15/issue-6/060506/Fiber-optic-nonlinear-endomicroscopy-with-focus-scanning-by-using-shape/10.1117/1.3523234.full}
\BIBentrySTDinterwordspacing

\bibitem{li_focus_2017}
\BIBentryALTinterwordspacing
A.~Li, W.~Liang, H.~Guan, Y.-T.~A. Gau, D.~E. Bergles, and X.~Li, ``\BIBforeignlanguage{EN}{Focus scanning with feedback-control for fiber-optic nonlinear endomicroscopy},'' \emph{\BIBforeignlanguage{EN}{Biomedical Optics Express}}, vol.~8, no.~5, pp. 2519--2527, May 2017, publisher: Optica Publishing Group. [Online]. Available: \url{https://opg.optica.org/boe/abstract.cfm?uri=boe-8-5-2519}
\BIBentrySTDinterwordspacing

\bibitem{lombardini_origin_2017}
\BIBentryALTinterwordspacing
A.~Lombardini, E.~R. Andresen, A.~Kudlinski, I.~Rimke, and H.~Rigneault, ``\BIBforeignlanguage{EN}{Origin and suppression of parasitic signals in {Kagomé} lattice hollow core fibers used for {SRS} microscopy and endoscopy},'' \emph{\BIBforeignlanguage{EN}{Optics Letters}}, vol.~42, no.~9, pp. 1824--1827, May 2017, publisher: Optica Publishing Group. [Online]. Available: \url{https://opg.optica.org/ol/abstract.cfm?uri=ol-42-9-1824}
\BIBentrySTDinterwordspacing

\end{thebibliography}


\end{document}